\documentclass[aps,preprint,superscriptaddress,nofootinbib]{revtex4}
\usepackage{graphicx,amsmath,amssymb}
\usepackage[usenames]{color}
\usepackage[colorlinks=true, urlcolor=navyblue, linkcolor=navyblue, citecolor=navyblue]{hyperref}
\definecolor{navyblue}{rgb}{0,0.08,0.45}

\newcommand{\mbf}[1]{\mathbf{#1}}
\newcommand{\half}{{\frac{1}{2}}}
\renewcommand{\bar}[1]{\overline{#1}}
\def\Dslash{\raise.15ex\hbox{/}\kern-.7em D}

\def\Journal#1#2#3#4{{#1} {{\bf #2},} {#4} {(#3)}}

\def\PRD{{Phys. Rev.} D}
\def\PRC{{Phys. Rev.} C}

\def\ZPC{{Z. Phys. C}}

\def\EPJC{{Eur. Phys. J.} C}

\def\MPLA{{Mod. Phys. Lett.} A}

\begin{document}
\preprint{SLAC-PUB-14525}

\title{AdS/QCD and Applications of Light-Front Holography}
\thanks{Presented by SJB and GdT at the Kavli Institute of Theoretical Physics (KITPC), Beijing, China, October 19 and 20, 2010. Review Article, to be published in {\it Communications in Theoretical Physics}.}
\author{Stanley J. Brodsky}
\affiliation
{SLAC National Accelerator Laboratory\\
Stanford University, Stanford, CA 94309, USA, and \\
CP$^3$-Origins,
Southern Denmark University, Odense, Denmark}
\thanks{Speaker}
\email{sjbth@slac.stanford.edu}

\author {Fu-Guang Cao}
\affiliation{Massey University, Palmerston North, New Zealand}
\email{f.g.cao@massey.ac.nz}

\author {Guy F. de T\'eramond}
\affiliation{Universidad de Costa Rica, San Jos\'e, Costa Rica}
\thanks{Speaker}
\email{gdt@asterix.crnet.cr}

\date{\today}

\begin{abstract}

{ Light-Front Holography  leads to a rigorous connection between hadronic amplitudes in a higher dimensional
 anti-de Sitter (AdS) space and frame-independent light-front wavefunctions of hadrons in $3+1$ physical space-time, thus providing a compelling physical interpretation of the AdS/CFT correspondence principle 
 and AdS/QCD, a useful framework which describes  the correspondence between theories in a modified AdS$_5$ background and confining field theories in physical space-time. 
 To a first semiclassical approximation, where quantum loops and quark masses are not included, this approach leads to  a single-variable light-front Schr\"odinger equation which 
 determines the eigenspectrum and the light-front wavefunctions of hadrons for general spin and orbital angular momentum. The coordinate $z$ in AdS space is uniquely identified 
 with  a Lorentz-invariant  coordinate $\zeta$ which measures the separation of the constituents within a hadron at equal light-front time.  The internal structure of hadrons is explicitly introduced
 and the angular momentum of the constituents plays a key role. We give an overview of the light-front holographic approach to strongly coupled QCD. 
 In particular, we study the photon-to-meson transition form factors (TFFs) $F_{M \gamma}(Q^2)$ for  $\gamma \gamma^* \to M$ using light-front holographic methods. 
 The results for the TFFs for the $\eta$ and $\eta^\prime$ mesons are also presented. Some novel features  of QCD are discussed, including the consequences of confinement for quark and gluon condensates. 
 A method for computing the hadronization of quark and gluon jets at the amplitude level is outlined.}

\end{abstract}

\maketitle

\section{Introduction to Light-Front Hamiltonian QCD}
\label{sec:Introduction}

A physically appealing and rigorous way to formulate quantum chromodynamics in both its nonperturbative and perturbative manifestations is Hamiltonian theory quantized on the light-front.
As Dirac~\cite{Dirac:1949cp} emphasized in 1949, quantizing a quantum field theory  at fixed light-front (LF) time $\tau = x^+ = x^0 + x^3$, the time marked by the
front of a light wave instead of the ordinary instant time $t = x^0$, provides a formalism which is explicitly frame-independent.   Note that when one takes a flash picture, 
the result is a measurement at a fixed time $\tau$ along the light-front. In practice, one always sets boundary conditions at fixed light-front time, not ordinary ``instant'' time 
since it requires the synchronization of many light-sources.

The QCD light-front Hamiltonian $H^{QCD} _{LF} =P^+ P^- - \mbf{P}^2_\perp$,  where $P^\pm = P^0 \pm P^3$, is constructed from the QCD Lagrangian using  standard methods of quantum field theory.~\cite{Brodsky:1997de} 
The commutators of operators at fixed $\tau$ are causal.  If one chooses the physical light-cone gauge $A^+= A^3 + A^0=0$, one can eliminate the dependent quark and gluon degrees of freedom.  
The resulting theory describes positive $k^+ > 0 $ quarks and gluons with physical polarization. In addition to the standard three- and four-point QCD interactions, the elimination of  the dependent fields leads
to new  four-point  interactions corresponding to  the exchange of instantaneous gluons  (analogous to the Coulomb interaction in Coulomb gauge) and instantaneous quark exchange (analogous to seagull interactions). 
The interactions conserve the plus  $\sum_i k^+_i$ and transverse momenta $\sum_i \mbf{k}_{\perp i}$  at every vertex, as well as the total angular momentum projection
$J^z =   \sum_{i = 1}^nS^z_i + \sum_{i=1}^{n-1} L^z_i$.  Each particle is on the positive energy mass shell in LF Hamiltonian theory $k^2 = k^+ k^- - \mbf{k}^2_\perp = m^2$ with $k^\pm = k^0 \pm k^3$. 
Since all quanta have positive $k^+$, the QCD
vacua are essentially trivial in the LF formalism - i.e.,  no vacuum processes appear, and thus the vacuum state is  identical to the free vacuum. The LF vacuum is already
normally-ordered, and one does not need to divide the S-matrix by vacuum loops.
 
A physical hadron in four-dimensional Minkowski space  with four-momentum $P_\mu$ and invariant
hadronic mass $M_H$ is an eigenstate of the light-front
Lorentz-invariant Hamiltonian equation for the relativistic bound-state system
$H^{QCD}_{LF} \vert  \psi(P) \rangle= P_\mu P^\mu \vert  \psi(P) \rangle = M_H^2 \vert  \psi(P) \rangle$,
where the light-front time evolution operator $P^-$ is determined canonically from the QCD Lagrangian.~\cite{Brodsky:1997de} In principle, the eigensolutions of $H^{QCD} _{LF}$
provide the complete bound-state and continuum scattering solutions necessary to describe hadron physics. In fact this has been performed explicitly for $QCD(1+1)$ for quark quanta with  arbitrary masses, flavors, 
 and for  finite numbers of colors $N_C$ using the discretized light-cone quantization (DLCQ)
method.~\cite{Hiller:2000pf}  Since Lorentz boosts are kinematical in the front form,~\cite{Dirac:1949cp}  the light-front wavefunctions of hadrons -- the hadronic  quark and gluon bound-state
eigensolutions of the QCD light-front Hamiltonian, are independent of the hadron's 3-momentum
$P^+$ and $\mbf{P}_\perp.$

One of the most important theoretical tools in atomic physics is the
Schr\"odinger wavefunction, which describes the quantum-mechanical
structure of  an atomic system at the amplitude level.  Light-front
wavefunctions (LFWFs) play a similar role in quantum chromodynamics,
providing a fundamental description of the structure and
internal dynamics of hadrons in terms of their constituent quarks
and gluons.
The LFWFs  $\psi_H(x_i, \mbf{k}_{\perp i}, \lambda_i)$ are the projections of the hadronic eigensolutions
$\vert  \psi(P) \rangle$ on the eigenstates of the free Hamiltonian,
where $x_i = k^+_i /P^+ $, $\sum_{i=1}^n k^+_i = P^+,$ $\sum_{i=1}^n \mbf{k}_{\perp i} =0$.
The state with the minimum number of constituents is referred to as the valence Fock state.
The LFWFs of bound states in QCD are thus
relativistic generalizations of the Schr\"odinger wavefunctions of
atomic physics, but they are determined at fixed light-front time
$\tau = x^0 + x^3$. The squares of the LFWFs summed over all Fock states give the generalized parton distributions of the hadrons. 
The structure functions measured in deep inelastic scattering  which satisfy  DGLAP evolution are derived from the squares of the LFWFs integrated over all variables 
but the struck quark's light-front momentum fraction $x = x_{bj}$.  The integral of the valence LFWF over transverse momentum squared up to $Q^2$ defines the distribution amplitude $\phi(x,Q^2)$, 
which controls exclusive processes at high $Q^2$. The logarithmic evolution of the distribution amplitude satisfies the ERBL evolution equation.

The simple structure of the LF vacuum allows an unambiguous
definition of the partonic content of a hadron in QCD and of hadronic light-front wavefunctions,
which relate its quark
and gluon degrees of freedom to their asymptotic hadronic state. The constituent spin and orbital angular momentum properties of the hadrons are also encoded in the LFWFs.  
As first noted by Casher and Susskind,~\cite{Casher:1974xd} chiral symmetry and other effects usually associated with the QCD vacuum are encoded within the Fock states of the hadrons.  
The implications of this for cosmology are discussed in
Sec. \ref{VacuumEffects}.

We have recently shown a remarkable
connection between the description of hadronic modes in AdS space and
the Hamiltonian formulation of QCD in physical space-time quantized
on the light-front at equal light-front time  $\tau$.~\cite{deTeramond:2008ht}
This connection, called ``Light-Front Holography"  is one of the most remarkable features of AdS/CFT.~\cite{Maldacena:1997re}  It  allows one to project the functional dependence
of the wavefunction $\Phi(z)$ computed  in the  AdS fifth dimension to the  hadronic frame-independent light-front wavefunction $\psi(x_i, \mbf{b}_{\perp i})$ in $3+1$ physical space-time. The
variable $z $ maps  to the LF variable $ \zeta(x_i, \mbf{b}_{\perp i})$.

On AdS space the physical  states are
represented by normalizable modes $\Phi_P(x, z) = e^{-iP \cdot x} \Phi(z)$,
with plane waves along Minkowski coordinates $x^\mu$ and a profile function $\Phi(z)$
along the holographic coordinate $z$. The hadronic invariant mass
$P_\mu P^\mu = M^2$  is found by solving the eigenvalue problem for the
AdS wave equation. Each  light-front hadronic state $\vert \psi(P) \rangle$ is dual to a normalizable string mode $\Phi_P(x,z)$.
For fields near the AdS boundary the behavior of $\Phi(z)$
depends on the scaling dimension of the corresponding interpolating operators.  Thus each hadron is identified by the twist of its interpolating operator at $z \to 0.$

The transverse coordinate $\zeta$ is closely related to the invariant mass squared  of the constituents in the LFWF  and its off-shellness  in  the LF kinetic energy, 
 and it is thus the natural variable to characterize the hadronic wavefunction.  In fact $\zeta$ is the only variable to appear
in the relativistic light-front Schr\"odinger equations predicted from AdS/QCD in the limit of zero quark masses.

We have shown that there exists a precise correspondence between the matrix elements of the electromagnetic current and the energy-momentum tensor of the fundamental hadronic constituents in QCD,
 with the corresponding transition amplitudes describing the interaction of string modes in anti-de Sitter space with the external sources which propagate in the AdS interior.  The agreement of the results for both
electromagnetic~\cite{Brodsky:2006uqa, Brodsky:2007hb} and gravitational~\cite{Brodsky:2008pf} hadronic transition amplitudes provides an important consistency test and verification of holographic mapping 
from AdS to physical observables defined on the light-front.
We have also studied the photon-to-meson transition form factors (TFFs) $F_{M \gamma}(Q^2)$ measured in $\gamma \gamma^* \to M$  reactions using light-front holographic
methods,~\cite{Brodsky:2011xx} processes which have been of intense experimental and theoretical interest. We review this recent work in Sec. \ref{PhotonMesonTFF}.

Conversely, one may take  the LF bound state Hamiltonian equation of motion in QCD as a starting point to  derive  relativistic wave equations in terms of an invariant transverse variable $\zeta$ which measures the
separation of the quark and gluonic constituents within the hadron
at the same LF time. The result is a single-variable light-front relativistic
Schr\"odinger equation,  which is
equivalent to the equations of motion which describe the propagation of spin-$J$ modes in a fixed  gravitational background asymptotic to AdS space. 
 Its eigenvalues give the hadronic spectrum and its eigenmodes represent the probability distribution of the hadronic constituents at a given scale.  Remarkably, the AdS equations correspond to 
 the kinetic energy terms of  the partons inside a hadron, whereas the interaction terms build confinement and
correspond to the truncation of AdS space in an effective dual gravity  approximation.~\cite{deTeramond:2008ht}
The identification of orbital angular momentum of the constituents is a key element in our description of the internal structure of hadrons using holographic principles,
since hadrons with the same quark content, but different orbital angular momenta, have different masses.

\section{Advantages of Evaluating Hadron Dynamics on the Light-Front}

As we have emphasized in Sec. \ref{sec:Introduction},
a remarkable feature of LFWFs is the fact that they are frame
independent; i.e., the form of the LFWF is independent of the
hadron's total momentum $P^+ = P^0 + P^3$ and $\mbf{P}_\perp.$
The simplicity of Lorentz boosts of LFWFs contrasts dramatically with the complexity of the boost of wavefunctions defined at fixed time $t.$~\cite{Brodsky:1968ea}
Light-front quantization is thus the ideal framework to describe the
structure of hadrons in terms of their quark and gluon degrees of freedom.  The
constituent spin and orbital angular momentum properties of the
hadrons are also encoded in the LFWFs.
The total  angular momentum projection~\cite{Brodsky:2000ii}
$J^z = \sum_{i=1}^n  S^z_i + \sum_{i=1}^{n-1} L^z_i$
is conserved Fock-state by Fock-state and by every interaction in the LF Hamiltonian.

Other advantageous features of light-front quantization include:

\begin{enumerate}

\item
If one quantizes QCD in the physical light-cone gauge (LCG) $A^+ =0$, then gluons  have physical angular momentum projections $S^z= \pm 1$. The orbital angular momenta of quarks and gluons are defined unambiguously, and there are no ghosts.   The empirical observation that quarks carry only a small fraction of the nucleon angular momentum highlights the importance of quark orbital angular momentum.  In fact the nucleon anomalous moment and the Pauli form factor are zero unless the quarks carry nonzero $L^z$.

\item
The gauge-invariant distribution amplitude $\phi(x,Q)$  is the integral of the valence LFWF in LCG integrated over the internal transverse momentum $k^2_\perp < Q^2$, because the Wilson line is trivial in this gauge. It is also possible to quantize QCD in  Feynman gauge in the light front.~\cite{Srivastava:1999gi}

\item
LF Hamiltonian perturbation theory provides a simple method for deriving analytic forms for the analog of Parke-Taylor amplitudes,~\cite{Motyka:2009gi} where each particle spin $S^z$ is quantized in the LF $z$ direction.  The gluonic $g^6$ amplitude  $T(-1 -1 \to +1 +1 +1 +1 +1 +1)$  requires $\Delta L^z =8;$ it thus must vanish at tree level since each three-gluon vertex has  $\Delta L^z = \pm 1.$ However, the order $g^8$ one-loop amplitude can be nonzero.

\item
Amplitudes in light-front perturbation theory are automatically renormalized using the ``alternate denominator'' subtraction method.~\cite{Brodsky:1973kb}  The application to QED has been checked at one and two loops.~\cite{Brodsky:1973kb}

\item
One can easily show using LF quantization that the anomalous gravitomagnetic moment $B(0)$  of a nucleon, as  defined from the spin flip matrix element of the energy-momentum tensor, vanishes Fock-state by Fock state,~\cite{Brodsky:2000ii} as required by the equivalence principle.~\cite{Teryaev:1999su}

\item
LFWFs obey the cluster decomposition theorem, providing the only proof of this theorem for relativistic bound states.~\cite{Brodsky:1985gs}

\item
The LF Hamiltonian can be diagonalized using the DLCQmethod.~\cite{Pauli:1985ps} This nonperturbative method is particularly elegant and useful for solving low-dimension quantum field theories such as
QCD$(1+1).$~\cite{Hornbostel:1988fb}

\item
LF quantization provides a distinction between static  (square of LFWFs) distributions versus non-universal dynamic structure functions,  such as the Sivers single-spin correlation and diffractive deep inelastic scattering which involve final state interactions.  The origin of nuclear shadowing and process independent anti-shadowing also becomes explicit.   This is discussed further in Sec. \ref{rescat}.

\item
LF quantization provides a simple method to implement jet hadronization at the amplitude level.  This is discussed in Sec.  \ref{hadronization}.

\item
The instantaneous fermion interaction in LF  quantization provides a simple derivation of the $J=0$
fixed pole contribution to deeply virtual Compton scattering.~\cite{Brodsky:2009bp}

\item
Unlike instant-time quantization, the Hamiltonian equation of motion in the LF is frame independent. This makes a direct connection of QCD with AdS/CFT methods possible.~\cite{deTeramond:2008ht}

\item
In the LF formalism, the effects usually associated with chiral and gluonic condensates are properties of the higher Fock states,~\cite{Casher:1974xd,Brodsky:2009zd} rather than the vacuum.  In the case of the Higgs model, the effect of the usual Higgs vacuum expectation value is replaced by a constant $k^+=0$ zero mode field.~\cite{Srivastava:2002mw}
\end{enumerate}

\section{Light-Front Holography\label{sec:LFH}}

A form factor in QCD is defined by the transition matrix element of a local quark current between hadronic states.  In AdS space form factors are computed from the overlap integral of normalizable modes with boundary currents which propagate in AdS space. The AdS/CFT duality incorporates the connection between the twist scaling dimension of the  QCD boundary interpolating operators
to the falloff of the
normalizable modes in AdS near its conformal boundary. If both quantities represent the same physical observable for any value of  the transferred momentum squared $q^2$,
a precise correspondence can be established between the string modes $\Phi$ in AdS space and the light front wavefunctions of hadrons $\psi_{n/H}$ in physical four dimensional space-time.~\cite{Brodsky:2006uqa} The same results follow from comparing the relativistic light-front Hamiltonian equation describing bound states in QCD with the wave equations describing the propagation of modes in a warped AdS
space.~\cite{deTeramond:2008ht}  In fact, one can systematically reduce  the LF  Hamiltonian equation to an effective relativistic wave equation, analogous to the AdS equations, by observing
that each $n$-particle Fock state has an essential dependence on the invariant mass of the system
and thus, to a first approximation, LF dynamics depend only on the invariant mass of the system.
In  impact space the relevant variable is a boost-invariant  variable $\zeta$
which measures the separation of the constituents at equal LF time.

\subsection{Electromagnetic Form Factor \label{EMFF}}

Light-Front Holography can be derived by observing the correspondence between matrix elements obtained in AdS/CFT with the corresponding formula using the light-front
representation.~\cite{Brodsky:2006uqa}
In the higher dimensional gravity theory, the hadronic matrix element  corresponds to
the  non-local coupling of an external electromagnetic field $A^M(x,z)$  propagating in AdS with the extended mode $\Phi(x,z)$~\cite{Polchinski:2002jw}
 \begin{multline} \label{FFAdSLF}
 \int d^4x \, dz  \, A^{M}(x,z)
 \Phi^*_{P'}(x,z) \overleftrightarrow\partial_M \Phi_P(x,z)
 \\ \sim
 (2 \pi)^4 \delta^4 \left( P'  \! - P - q\right) \epsilon_\mu  \langle \psi(P')  \vert J^\mu \vert \psi(P) \rangle ,
 \end{multline}
 where the coordinates of AdS$_5$ are the Minkowski coordinates $x^\mu$ and $z$ labeled $x^M = (x^\mu, z)$, with
$M = 1, \cdots 5$,
 and $g$ is the determinant of the metric tensor. The expression on the right-hand side  represents the QCD EM transition amplitude in physical space-time. It is the EM matrix element of the quark current  $J^\mu = e_q \bar q \gamma^\mu q$, and represents a local coupling to pointlike constituents. Although the expressions for the transition amplitudes look very different, one can show  that a precise mapping of the $J^+$ elements  can be carried out at fixed light-front time.

The light-front electromagnetic form factor in impact
space~\cite{Brodsky:2006uqa,Brodsky:2007hb,Soper:1976jc} can be written as a sum of overlap of light-front wave functions of the $j = 1,2, \cdots, n-1$ spectator
constituents:
\begin{equation} \label{eq:FFb}
F(q^2) =  \sum_n  \prod_{j=1}^{n-1}\int d x_j d^2 \mbf{b}_{\perp j}   \sum_q e_q
            \exp \! {\Bigl(i \mbf{q}_\perp \! \cdot \sum_{j=1}^{n-1} x_j \mbf{b}_{\perp j}\Bigr)}
 \left\vert  \psi_{n/H}(x_j, \mbf{b}_{\perp j})\right\vert^2 ,
\end{equation}
where the normalization is defined by
\begin{equation}  \label{eq:Normb}
\sum_n  \prod_{j=1}^{n-1} \int d x_j d^2 \mathbf{b}_{\perp j}
\vert \psi_{n/H}(x_j, \mathbf{b}_{\perp j})\vert^2 = 1.
\end{equation}

The formula  is exact if the sum is over all Fock states $n$.~\cite{Drell:1969km, West:1970av}
For definiteness we shall consider
the $\pi^+$  valence Fock state
$\vert u \bar d\rangle$ with charges $e_u = \frac{2}{3}$ and $e_{\bar d} = \frac{1}{3}$.
For $n=2$, there are two terms which contribute to the $q$-sum in (\ref{eq:FFb}).
Exchanging $x \leftrightarrow 1 \! - \! x$ in the second integral  we find
\begin{equation}  \label{eq:PiFFb}
 F_{\pi^+}(q^2)  =  2 \pi \int_0^1 \! \frac{dx}{x(1-x)}  \int \zeta d \zeta \,
J_0 \! \left(\! \zeta q \sqrt{\frac{1-x}{x}}\right)
\left\vert \psi_{u \bar d/ \pi}\!(x,\zeta)\right\vert^2,
\end{equation}
where $\zeta^2 =  x(1  -  x) \mathbf{b}_\perp^2$ and $F_{\pi^+}(q\!=\!0)=1$.

We now compare this result with the electromagnetic (EM) form-factor
in  AdS  space time. 
The incoming electromagnetic field propagates in AdS according to
$A_\mu(x^\mu ,z) = \epsilon_\mu(q) e^{-i q \cdot x} V(q^2, z)$,
where $V(q^2,z)$, the bulk-to-boundary propagator, is the solution of the AdS wave equation
with boundary conditions $V(q^2 = 0, z ) = V(q^2, z = 0) = 1$.~\cite{Polchinski:2002jw}
The propagation of the pion in AdS space is described by a normalizable mode
$\Phi_P(x^\mu, z) = e^{-i P  \cdot x} \Phi(z)$ with invariant  mass $P_\mu P^\mu = \mathcal{M}_\pi^2$ and plane waves along Minkowski coordinates $x^\mu$.  
Factoring out the plane wave dependence of the AdS fields  we find
the transition amplitude   $(Q^2 = - q^2 >0$)
\begin{equation}
\langle P' \vert J^\mu \vert P \rangle = \left(P + P' \right)^\mu F(Q^2),
\end{equation}
where we have extracted the overall factor  $ (2 \pi)^4 \delta^4 \left( P'  \! - P - q\right)$ from momentum
conservation at the vertex from integration over Minkowski variables in (\ref{FFAdSLF}).  We find for $F(Q^2)$~\cite{Polchinski:2002jw}
\begin{equation}
F(Q^2) = R^3 \int \frac{dz}{z^3} \, V(Q^2, z) \vert \Phi(z) \vert^2,
\label{eq:FFAdS}
\end{equation}
where $F(Q^2 = 0) = 1$. 
Using the integral representation of   $V(Q^2, z)$ 
\begin{equation} \label{eq:intJ}
V(Q^2, z) =  z Q K_1(z Q)=  \int_0^1 \! dx \, J_0\negthinspace \left(\negthinspace\zeta Q
\sqrt{\frac{1-x}{x}}\right) ,
\end{equation}
 we write the AdS electromagnetic form-factor as
\begin{equation}
F(Q^2)  =    R^3 \! \int_0^1 \! dx  \! \int \frac{dz}{z^3} \,
J_0\!\left(\!z Q\sqrt{\frac{1-x}{x}}\right) \left \vert\Phi(z) \right\vert^2 .
\label{eq:AdSFx}
\end{equation}
To compare with  the light-front QCD  form factor expression (\ref{eq:PiFFb})  we 
write the LFWF as 
\begin{equation} \label{psiphi}
\psi(x,\zeta, \varphi) = e^{i M \varphi} X(x) \frac{\phi(\zeta)}{\sqrt{2 \pi \zeta}} ,
\end{equation}
thus factoring out the angular dependence $\varphi$ in the transverse LF plane, the longitudinal $X(x)$ and
transverse mode $\phi(\zeta)$. The factorization of the LFWF given by (\ref{psiphi}) is a natural factorization in the light front formalism since the
corresponding canonical generators, the longitudinal and transverse generators $P^+$ and $\mbf{P}_\perp$ and the $z$-component of the orbital angular momentum
$J^z$, are kinematical generators which commute with the LF Hamiltonian generator $P^-$.~\cite{Dirac:1949cp} If both expressions for the form factor are identical for arbitrary values of $Q$,
we obtain $\phi(\zeta) = (\zeta/R)^{3/2} \Phi(\zeta)$ and $X(x) = \sqrt{x(1-x)}$,~\cite{Brodsky:2006uqa}
where we identify the transverse impact LF variable $\zeta$ with the holographic variable $z$,
$z \to \zeta = \sqrt{x(1-x)} \vert \mbf b_\perp \vert$.
We choose the normalization
 $ \langle\phi\vert\phi\rangle = \int \! d \zeta \,
 \vert \langle \zeta \vert \phi\rangle\vert^2 = 1$.

Extension of the results to arbitrary $n$ follows from the $x$-weighted definition of the
transverse impact variable of the $n-1$ spectator system:~\cite{Brodsky:2006uqa}
\begin{equation} \label{zeta}
\zeta = \sqrt{\frac{x}{1-x}} ~ \Big\vert \sum_{j=1}^{n-1} x_j \mbf{b}_{\perp j} \Big\vert ,
\end{equation}
where $x = x_n$ is the longitudinal
momentum fraction of the active quark.
A recent application of the light-front holographic ideas
has been used
to compute the helicity-independent generalized parton distributions (GPDs) of quarks
in a  nucleon in the zero skewness case.~\cite{Vega:2010ns}

Conserved currents are not renormalized and correspond to five dimensional massless fields propagating in AdS according to the relation
$(\mu R)^2 = (\Delta - p) (\Delta + p -  4)$  for a $p$ form in $d=4$. In the usual AdS/QCD framework~\cite{Erlich:2005qh, DaRold:2005zs} this  corresponds to $\Delta = 3$ or 1, the canonical dimensions of
an EM current and the massless gauge field respectively.  Normally one uses a hadronic  interpolating operator  with minimum twist $\tau$ to identify a hadron in AdS/QCD and to predict the power-law fall-off behavior of its form factors and other hard
scattering amplitudes;~\cite{Polchinski:2001tt}  e.g.,  for a two-parton bound state $\tau = 2$.   However, in the case of a current, one needs to  use  an effective field operator  with dimension $\Delta =3.$ The apparent inconsistency between twist and dimension is removed by noticing that in the light-front one chooses to calculate the  matrix element of the twist-3 plus  component of the current  $J^+$,~\cite{Brodsky:2006uqa, Brodsky:2007hb} in order to avoid coupling to Fock states with different numbers of constituents.

\begin{figure}[h]
\includegraphics[angle=0,width=8.6cm]{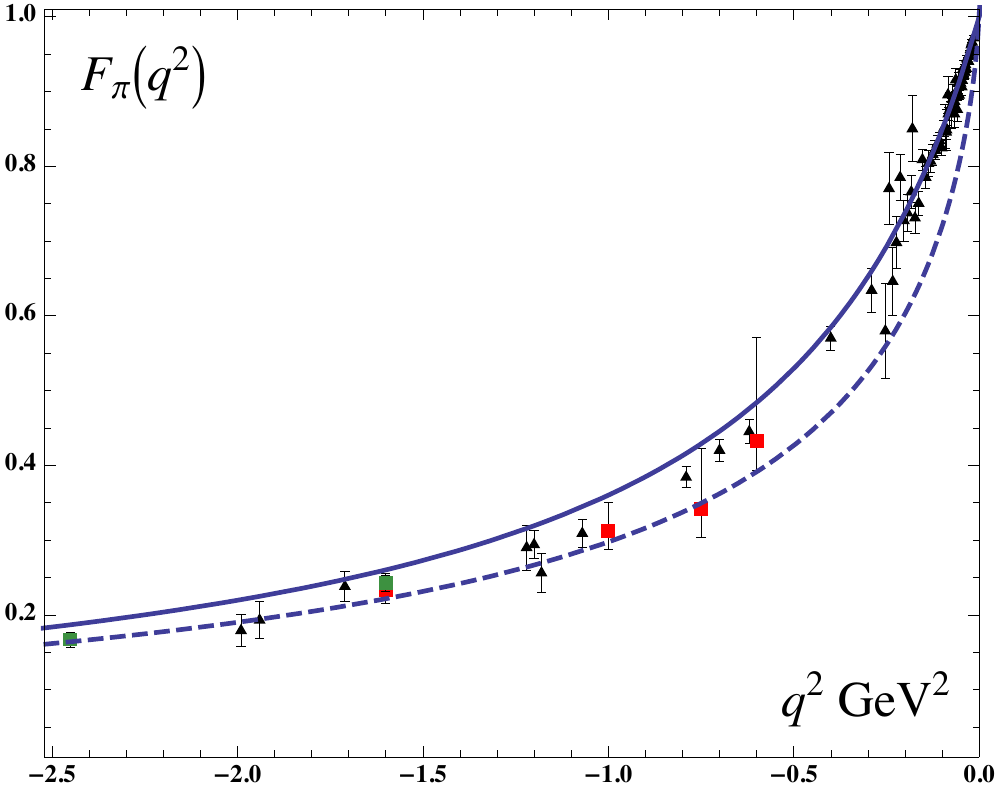}
\caption{ \small Space-like pion form factor $F_\pi(q^2)$.  Continuous line: confined current; dashed  line: free current.
Triangles are the data compilation  from Baldini,~\cite{Baldini:1998qn}  boxes  are JLAB data.~\cite{Tadevosyan:2007yd, Horn:2006tm}}
 \label{PionFFSL}
\end{figure}

The results described above correspond to a ``free" current propagating on AdS space and dual to the EM pointlike current in the DYW light-front formula,  which allow us to map state-by-state.\footnote{In general the mapping relates the AdS density $\Phi^2(z)$ to an  effective LF single particle transverse density.~\cite {Brodsky:2006uqa}}
This mapping has the shortcoming that the pole structure of the form factor is not built on the timelike region. Furthermore, the moments of the form factor at $Q^2=0$ diverge, giving for example an infinite charge radius.  The pole structure is generated when the EM current is confined, this means, when the EM current propagates on
a IR deformed AdS space to mimic confinement. This also leads to finite moments at $Q^2=0$, as illustrated on Fig. \ref{PionFFSL}.

Hadronic form factors  for the harmonic potential $\kappa^2 z^2$  have a simple analytical form~\cite{Brodsky:2007hb} 
\begin{equation} \label{F}   
 F_\tau(Q^2) =  \frac{1}{{\Big(1 + \frac{Q^2}{M^2_\rho} \Big) }
 \Big(1 + \frac{Q^2}{M^2_{\rho'}}  \Big)  \cdots 
       \Big(1  + \frac{Q^2}{M^2_{\rho^{\tau-2}}} \Big)} ,
\end{equation}
which is  expressed as a $\tau - 1$ product of poles along the vector meson Regge radial trajectory.
For a pion, for example, the lowest Fock state -- the valence state -- is a twist-2 state, and thus the form factor is the well known monopole form.~\cite{Brodsky:2007hb}
The remarkable analytical form of (\ref{F}),
expressed in terms of the $\rho$ vector meson mass and its radial excitations, incorporates the correct scaling behavior from the constituent's hard scattering with the photon and the mass gap from confinement.  It is also apparent from (\ref{F}) that the higher-twist components in the Fock expansion are relevant for the computation of hadronic form factors, particularly for the time-like region which is particularly sensitive to the detailed structure of the amplitudes.~\cite{deTeramond:2010ez} For a confined EM current in AdS a precise mapping can also be carried out to the DYW expression for the form factor. In this case we we find an effective LFWF, which corresponds to a superposition of an infinite number of Fock states.~\cite{Brodsky:2011xx}

Light front holography provides a precise relation of the fifth-dimensional mass $\mu$ with the total  and orbital angular momentum of a hadron in the transverse LF plane $(\mu R)^2 = - (2 - J)^2 + L^2$, $L = \vert L^z\vert$,~\cite{deTeramond:2008ht} and thus  a conserved EM current  corresponds to poles along the $J=L=1$ radial  trajectory. For the twist-3 computation of the space-like form factor, which involves the current $J^+$, the poles do not correspond to the physical poles of the twist-2 transverse current $\mbf{J}_\perp$ presented
in the annihilation channel, namely the $J=1$, $L=0$ radial trajectory. Consequently, the location of the poles in the final result should be shifted to their physical positions.~\cite{Brodsky:2011xx} When this is done, the results agree extremely well with the space-like pion form factor data as shown in Fig.  \ref{PionFFSL}, as well as for the space-like proton Dirac elastic and transition form factor data.~\cite{deTeramond:2011qp}  The non-perturbative effects from the dressed current correspond to an infinite sum of diagrams. One should however be careful to avoid a double counting of terms.

\subsection{Gravitational Form Factor}

Matrix elements of the energy-momentum tensor $\Theta^{\mu \nu} $ which define the gravitational form factors play an important role in hadron physics.  Since one can define $\Theta^{\mu \nu}$ for each parton, one can identify the momentum fraction and  contribution to the orbital angular momentum of each quark flavor and gluon of a hadron. For example, the spin-flip form factor $B(q^2)$, which is the analog of the Pauli form factor $F_2(Q^2)$ of a nucleon, provides a  measure of the orbital angular momentum carried by each quark and gluon constituent of a hadron at $q^2=0.$   Similarly,  the spin-conserving form factor $A(q^2)$, the analog of the Dirac form factor $F_1(q^2)$, allows one to measure the momentum  fractions carried by each constituent.
This is the underlying physics of Ji's sum rule:~\cite{Ji:1996ek}
$\langle J^z\rangle = \half [ A(0) + B(0)] $,  which has prompted much of the current interest in
the GPDs measured in deeply
virtual Compton scattering.  An important constraint is $B(0) = \sum_i B_i(0) = 0$;   i.e., the anomalous gravitomagnetic moment of a hadron vanishes when summed over all the constituents $i$. This was originally derived from the equivalence principle of gravity.~\cite{Teryaev:1999su}  The explicit verification of these relations, Fock state by Fock state, can be obtained in the LF quantization of QCD in  light-cone
gauge.~\cite{Brodsky:2000ii} Physically $B(0) =0$ corresponds to the fact that the sum of the $n$ orbital angular momenta $L$ in an $n$-parton Fock state must vanish since there are only $n-1$ independent orbital angular momenta.

The matrix element of the energy-momentum tensor  for the hadronic transition
$P \to P'$,
follows from  the
coupling of the hadronic mode $\Phi_P(x,z)$  with the external graviton field propagating in AdS
space~\cite{Abidin:2008ku}
\begin{multline}
 \int \! d^4x \, dz \sqrt{g}\, h_{\ell m}  \left(
\partial^{\ell} \Phi_{P'}^* \partial^{m} \Phi_P+
\partial^{m} \Phi_{P'}^* \partial^{\ell} \Phi_P \right)
 \\ \sim
 (2 \pi)^4 \delta^4 \left( P'  \! - P - q\right) \epsilon_{\mu \nu}  \langle \psi(P')  \vert \Theta^{\mu \nu} \vert \psi(P) \rangle.
\label{eq:T}
\end{multline}
The expression on the right-hand side  is the QCD matrix elements of the energy-momentum tensor in physical space-time
\begin{equation} \label{eq:emt}
\Theta_{\mu \nu} =  \frac{1}{2}
\bar \psi  i \! \left( \gamma_\mu D_\nu + \gamma_\nu D_\mu \right) \psi
- {\rm g}_{\mu \nu} \bar \psi \left( i \Dslash - m\right)\psi
 - G^a_{\mu \lambda} {G^a_\nu}^{\hspace{0.5pt} \lambda}
 + \tfrac{1}{4} {\rm g}_{\mu\nu}  G^a_{\lambda \sigma} G^{a \hspace{1pt} \lambda \sigma},
\end{equation}
 and represents a local coupling to pointlike constituents. As for the EM form factor,  the expressions for the transition amplitudes look very different, but one can show  that a precise mapping of the $\Theta^{++}$ elements  can be carried out at fixed light-front time.

The LF expression for the helicity-conserving gravitational form factor in impact space
is~\cite{Brodsky:2008pf}
\begin{equation} \label{eq:Ab}
A(q^2) =  \sum_n  \prod_{j=1}^{n-1}\int d x_j d^2 \mbf{b}_{\perp j}  \sum_f  x_f
\exp \! {\Bigl(i \mbf{q}_\perp \! \cdot \sum_{j=1}^{n-1} x_j \mbf{b}_{\perp j}\Bigr)}
\left\vert  \psi_{n/H}(x_j, \mbf{b}_{\perp j})\right\vert^2,
\end{equation}
which includes the contribution of each struck parton with longitudinal momentum $x_f$
and corresponds to a change of transverse momentum
$x_j \mbf{q}_\perp$
for each of the $j = 1, 2, \cdots, n-1$ spectators.
For $n=2$, there are two terms which contribute to the $f$-sum in  (\ref{eq:Ab}).
Exchanging $x \leftrightarrow 1-x$ in the second integral we find
\begin{equation} \label{eq:PiGFFb}
A_{\pi}(q^2) =  4 \pi \int_0^1 \frac{dx}{(1-x)}  \int \zeta d \zeta \,
J_0 \! \left(\! \zeta q \sqrt{\frac{1-x}{x}}\right)
\left\vert \psi_{q \bar q/ \pi}\!(x,\zeta)\right\vert^2,
\end{equation}
where $\zeta^2 =  x(1-x) \mathbf{b}_\perp^2$ and  $A_{\pi}(0) = 1$.

 We now consider the expression for the hadronic gravitational form factor in AdS space, which is obtained by perturbing the metric from the static
 AdS geometry~\cite{Abidin:2008ku}
 \begin{equation} \label{eq:AdSz}
ds^2 = \frac{R^2}{z^2} \left(\eta_{\mu \nu} dx^\mu dx^\nu - dz^2\right) .
\end{equation}
Factoring out the plane wave dependence of the AdS fields   we find
the transition amplitude
\begin{equation}
\left\langle P' \left\vert \Theta_\mu^{\, \nu} \right\vert P \right\rangle
=  \left( P^\nu P'_\mu + P_\mu  P'^\nu \right) A(Q^2),
\end{equation}
where we have extracted the overall factor  $ (2 \pi)^4 \delta^4 \left( P'  \! - P - q\right)$ from momentum
conservation at the vertex from integration over Minkowski variables in (\ref{eq:T}).  We find for $A(Q^2)$~\cite{Abidin:2008ku}
 \begin{equation}
A(Q^2)  =  R^3 \! \! \int \frac{dz}{z^3} \, H(Q^2, z) \left\vert\Phi_\pi(z) \right\vert^2,
\end{equation}
where $A(Q^2 = 0) =1$ and $H(Q^2, z) = \half  Q^2 z^2  K_2(z Q)$.
Using the integral representation of $H(Q^2,z)$
\begin{equation} \label{eq:intHz}
H(Q^2, z) =  2  \int_0^1\!  x \, dx \, J_0\!\left(\!z Q\sqrt{\frac{1-x}{x}}\right) ,
\end{equation}
we can write the AdS gravitational form factor
\begin{equation}
A(Q^2)  =  2  R^3 \! \int_0^1 \! x \, dx  \! \int \frac{dz}{z^3}  \,
J_0\!\left(\!z Q\sqrt{\frac{1-x}{x}}\right) \left \vert\Phi(z) \right\vert^2 .
\label{eq:AdSAx}
\end{equation}
Comparing with the QCD  gravitational form factor (\ref{eq:PiGFFb}) for arbitrary values of $Q$ we find an  identical  relation between the LF wave function $\psi(x,\zeta)$ and the AdS wavefunction $\Phi(z)$ and the factorization given
 by Eq. (\ref{psiphi}), which was obtained in Sect. \ref{EMFF} from the mapping of the pion electromagnetic transition amplitude.

As for the case of the electromagnetic form factor, the AdS mapping of the gravitational form factor is carried out in light-front holography for the plus-plus components of the energy-momentum tensor $\Theta^{++}$. The twist of this operator is $\tau = 4$ and coincides with the canonical conformal dimension of the energy-momentum tensor.

\section{Photon-to-meson transition form factors\label{PhotonMesonTFF}}

The light-front holographic methods described in Sec. \ref{sec:LFH}
can be used in the study of other exclusive processes. In this section we review such an application in the analysis of
the two-photon processes $\gamma\gamma \rightarrow M$ with $M$ being a pseudoscalar meson. \cite{Brodsky:2011xx} The pion transition form factor between a photon and pion measured in the $e^- e^-\to e^- e^- \pi^0$  process, with one tagged electron, is the simplest bound-state process in QCD.
It can be predicted from first principles in the asymptotic $Q^2 \to \infty$  limit.~\cite{Lepage:1980fj}  More generally,
the pion TFF at large $Q^2$ can be calculated at leading twist as a convolution of a perturbative hard scattering amplitude $T_H(\gamma \gamma^* \to q \bar q)$
and a gauge-invariant meson distribution amplitude (DA) which incorporates the nonperturbative dynamics of the QCD bound-state.~\cite{Lepage:1980fj}

The BaBar\ Collaboration has reported measurements of the
transition form factors from $\gamma^* \gamma \to M$ process for the $\pi^0$,~\cite{Aubert:2009mc}
$\eta$, and $\eta^\prime$~\cite{BaBar_eta, Druzhinin:2010bg} pseudoscalar mesons for a momentum   transfer  range much larger than
previous measurements.~\cite{CELLO,CLEO}  Surprisingly, the BaBar\ data for the $\pi^0$-$\gamma$ TFF
exhibit a rapid growth for $Q^2 > 15$ GeV$^2$, which is unexpected from QCD predictions. In contrast, the data for  the  $\eta$-$\gamma$ and  $\eta'$-$\gamma$
TFFs are in agreement with previous experiments and theoretical predictions.
Many theoretical studies have been devoted to explaining BaBar's experimental results.
\cite{BaBar_expln_LiM09, MikhailovS09,WuH10,BaBar_Expln_BroniowshiA10,RobertsRBGT10,PhamP11,Kroll10,GorchetinGS11,ZuoJH10,ADorokhov10,SAgaevBOP11,Cao:2011,BakulevMPS11}

\subsection{The Chern-Simons Structure of the Meson Transition Form Factor in AdS Space \label{sec:CSStructure}}

To describe the pion transition form factor within the framework of holographic QCD we need to explore the mathematical structure of higher-dimensional forms 
in the five dimensional action, since the amplitude (\ref{FFAdSLF}) can only account for the elastic form factor $F_M(Q^2)$.
For example, in the five-dimensional compactification of Type II B supergravity~\cite{Pernici:1985ju, Gunaydin:1985cu}
there is a Chern-Simons term in the action in addition to the usual Yang-Mills term $F^2$.~\cite{Witten:1998qj}
In the case of the $U(1)$ gauge theory the CS action is of the form $\epsilon^{L M N P Q} A_L \partial_M A_N \partial_P A_Q$ in the five
dimensional Lagrangian.~\cite{Hill:2006ei} 
The CS action is not gauge invariant: under a gauge transformation it changes by a total derivative which gives a surface term.   

The Chern-Simons form is the product of three fields at the same point in five-dimensional space corresponding to a local interaction.  Indeed the five-dimensional CS action is responsible for the anomalous coupling of mesons to photons and has been used to describe, for example, the $\omega \to \pi \gamma$~\cite{Pomarol:2008aa} decay as well as the 
$\gamma  \gamma^* \to \pi^0$~\cite{Grigoryan:2008up, Zuo:2011sk}
 and  $\gamma^* \rho^0 \to \pi^0$~\cite{Zuo:2009hz} processes.~\footnote{The anomalous EM couplings to mesons in the Sakai and Sugimoto model is described in Ref. \cite{Sakai:2005yt}.}

The hadronic matrix element for the anomalous electromagnetic coupling to mesons in the higher gravity theory is
given by the five-dimensional CS  amplitude
\begin{multline} \label{eq:TFFAdS1}
\int d^4 x \int dz \, \epsilon^{L M N P Q} A_L \partial_M A_N \partial_P A_Q  \\ \sim
(2 \pi)^4 \delta^{(4)} \left(P + q - k\right) F_{\pi \gamma}(q^2) \epsilon^{\mu \nu \rho \sigma} \epsilon_\mu(q) P_\nu \epsilon_\rho(k) q_\sigma,
\end{multline}
which includes the pion field as well as the external photon fields by identifying the fifth component of $A$ with the meson mode in AdS space.~\cite{Hill:2004uc}
In the r.h.s of (\ref{eq:TFFAdS1})  $q$ and $k$ are the momenta of the virtual and on-shell incoming photons respectively  with  corresponding polarization vectors  $\epsilon_\mu(q)$ 
and $\epsilon_\mu(k)$ for  the amplitude   $\gamma \gamma^* \to \pi^0$.
The momentum of the outgoing  pion is $P$.

The pion transition form factor $F_{\pi \gamma}(Q^2)$  can be computed from first principles in QCD. To leading 
leading order in $\alpha_s(Q^2)$ and leading twist the result is~\cite{Lepage:1980fj} ($Q^2 = - q^2 >0$)
\begin{equation}
Q^2 F_{\pi \gamma}(Q^2)=\frac{4}{\sqrt{3}} \int_0^1  {\rm d} x \frac{\phi(x,{\bar x} Q)}{\bar x}
\left[ 1+ O \left(\alpha_s,\frac{m^2}{Q^2} \right) \right],
\label{eq:TFLB1}
\end{equation}
where $x$ is the longitudinal momentum fraction of the quark struck by the virtual photon in the hard scattering process
and ${\bar x}=1-x$ is the longitudinal momentum fraction of the spectator quark.  
The pion distribution amplitude $\phi(x,Q)$ in the light-front formalism~\cite{Lepage:1980fj} is the integral of the 
valence $q \bar q$ LFWF in light-cone gauge $A^+=0$
\begin{equation}
\phi(x,Q)=\int_0^{Q^2}\frac{ d^2 \mbf{k}_\perp}{16 \pi^3} \psi_{q \bar q/ \pi}(x, \mbf{k}_\perp),
\label{eq:DALC}
\end{equation}
and has the asymptotic form~\cite{Lepage:1980fj}  $\phi(x, Q \to \infty) = \sqrt{3} f_\pi x (1-x)$; thus the  leading order  QCD result  for the TFF at the asymptotic limit
is obtained,~\cite{Lepage:1980fj}
\begin{equation} \label{TFFasy}
Q^2 F_{\pi \gamma}(Q^2 \rightarrow \infty)=2 f_\pi.
\end{equation} 

We now compare the QCD expression on the r.h.s. of  (\ref{eq:TFFAdS1}) with the AdS transition amplitude on the l.h.s. As for the elastic form factor discussed in Sec. \ref{EMFF}, the incoming off-shell photon is represented by the propagation of the non-normalizable electromagnetic solution  in AdS space,
$A_\mu(x^\mu ,z) = \epsilon_\mu(q) e^{-i q \cdot x} V(q^2, z)$,
where $V(q^2,z)$ is the bulk-to-boundary propagator with boundary conditions $V(q^2 = 0, z ) = V(q^2, z = 0) = 1$.~\cite{Polchinski:2002jw}
Since the incoming photon with momentum $k$ is on its mass shell, $k^2 = 0$,  its wave function is $A_\mu(x^\mu, z) = \epsilon_\mu(k) e^{ i k \cdot x}$.   
Likewise, the propagation of the pion in AdS space is described by a normalizable mode
$\Phi_{P}(x^\mu, z) = e^{-i P  \cdot x} \Phi_\pi(z)$ with invariant  mass $P_\mu P^\mu = \mathcal{M}_\pi^2 =0$   
in the chiral limit for massless quarks. 
The normalizable mode $\Phi(z)$ scales as $\Phi(z) \to z^{\tau=2}$  in the limit $z \to 0$, since the leading interpolating operator for the pion has twist two.
 A simple dimensional analysis implies that  $A_z \sim  \Phi_\pi(z)/ z$, matching the twist scaling dimensions: two for the pion and one for the EM field.  Substituting in
(\ref{eq:TFFAdS1}) the expression given above for the 
 the pion and the EM fields  propagating in  AdS,   and extracting the overall factor  $(2 \pi)^4 \delta^4 \left( P'  \! - q - k\right)$ upon integration over Minkowski variables in (\ref{FFAdSLF}) we find  $(Q^2 = - q^2 > 0)$
\begin{equation}  \label{eq:TFFAdS2}
F_{\pi \gamma}(Q^2) = \frac{1}{2 \pi} \int_0^\infty   \frac{d z}{z} \,  \Phi_\pi(z)  V\!\left(Q^2, z\right) ,
\end{equation}
where the normalization is fixed by the asymptotic QCD prediction (\ref{TFFasy}). We have defined our units such that the AdS radius $R=1$. 

Since the LF mapping of (\ref{eq:TFFAdS2}) to the asymptotic QCD prediction (\ref{TFFasy}) only depends on the asymptotic behavior near the boundary of AdS space, the result is independent of the particular model used to 
modify the large $z$ IR region of AdS space.  At large enough $Q$, the important contribution to (\ref{TFFasy}) only comes from the region near $z \sim 1/Q$ where 
$\Phi(z) = 2 \pi f_\pi z^2 + \mathcal{O}(z^4)$.  Using the integral
\begin{equation}
\int_0^\infty dx \, x^\alpha K_1(x) = 2^{\alpha -2} \alpha \, \left[\Gamma \! \left(\frac{\alpha}{2}\right)\right]^2,
~~~ {\rm Re}(\alpha) >1,
\end{equation}
we recover the asymptotic result  (\ref{TFFasy})
\begin{equation}
Q^2 F_{\pi \gamma}(Q^2 \rightarrow \infty)=2 f_\pi + \mathcal{O}\left(\frac{1}{Q^2}\right),
\end{equation} 
with the pion decay constant $f_\pi$
\begin{equation} \label{eq:fpi}
f_\pi = \frac{1}{4 \pi} \frac{\partial_z\Phi^\pi(z)}{z} \Big\vert_{z=0}.
\end{equation}
Since the pion field is identified as the fifth component of $A_M$,
the CS form $\epsilon^{L M N P Q} A_L \partial_M A_N \partial_P A_Q$ is similar in form to an axial current;  this correspondence can explain why the resulting pion distribution amplitude has the asymptotic form. 

In Ref. \cite{Grigoryan:2008up} the pion TFF was studied in the framework of a CS extended hard-wall AdS/QCD model with  $A_z \sim \partial_z \Phi(z)$.  The  expression for the TFF
which follows from (\ref{eq:TFFAdS1}) then vanishes at $Q^2 =0$, and has to be corrected by the introduction of a surface term at the IR wall.~\cite{Grigoryan:2008up}
However, this procedure is only possible for a model with a sharp cutoff.
The pion TFF has also been studied using the holographic approach to  QCD in Refs.~\cite{Cappiello:2010uy, Stoffers:2011xe}.

\subsection{A Simple Holographic Confining Model}
\label{sec:FreeCurrent}

QCD  predictions of the TFF  correspond to the local coupling of the free electromagnetic current to the elementary constituents in the interaction representation.~\cite{Lepage:1980fj} 
To compare with  QCD results, we first consider a simplified model where the non-normalizable mode $V(Q^2,z)$ for the  EM current satisfies the  ``free" AdS equation
subject to the boundary conditions $V(Q^2 =0, z) = V(Q^2, z =0) = 1$; thus the solution  $V(Q^2, z) = z Q K_1(z Q)$,  dual  to the free electromagnetic current.~\cite{Brodsky:2006uqa}
To describe the normalizable mode representing the pion we take the soft-wall exponential form,
\begin{equation}  \label{eq:Phitau}
\Phi^\tau(z) =   \sqrt{\frac{2 P_{\tau}}{\Gamma(\tau \! - \! 1)} } \, \kappa^{\tau -1} z ^{\tau} e^{- \kappa^2 z^2/2},
\end{equation} 
with normalization
\begin{equation} \label{eq:PhitauNorm}
\langle\Phi^\tau\vert\Phi^\tau\rangle = \int \frac{dz}{z^3} \, e^{- \kappa^2 z^2} \Phi^\tau(z)^2  = P_\tau,
\end{equation}
where $P_\tau$ is the probability for the twist $\tau$ mode (\ref{eq:Phitau}).
This agrees with the fact that the field $\Phi^\tau$ couples to a local hadronic interpolating operator of twist $\tau$ 
defined at the asymptotic boundary of AdS space, and thus
the scaling dimension of $\Phi^\tau$ is $\tau$. 

For twist $\tau=2$ we have
\begin{equation}   \label{eq:Phipi}
\Phi_{q \bar q/ \pi} (z) =   \sqrt{2 P_{q \bar q}} \, \kappa \, z^2 e^{- \kappa^2 z^2/2},
\end{equation} 
with normalization
\begin{equation}
\langle\Phi_{q \bar q/ \pi} \vert \Phi_{q \bar q/ \pi} \rangle 
= \int \frac{dz}{z^3} \, e^{- \kappa^2 z^2} \Phi_{q \bar q/ \pi}^2(z)  = P_{q \bar q},
\end{equation}
where $P_{q \bar q}$ is the probability for the valence state.
The LF mapping of (\ref{eq:Phipi}) has also a convenient exponential form and has been studied considerably in the literature.~\cite{Cao:2011}  
The exponential form of the LFWF  in momentum space has important support only when the virtual states are near the energy shell, and thus it  implements in a natural way the requirements of  the bound state dynamics.
From (\ref{eq:fpi}) the pion decay constant  is
\begin{equation} \label{eq:fpiPhi}
f_\pi = \sqrt{P_{q \bar q} } \, \frac{\kappa}{\sqrt{2} \pi}.
\end{equation}
 It is not possible in this model to introduce a surface term as in Ref. \cite{Grigoryan:2008up} to match the
value of the TFF at $Q^2 = 0$  derived from the decay $\pi^0 \to \gamma \gamma$. Instead, higher Fock components which modify the pion wave function at large distances are required 
to satisfy this low-energy constraint naturally. Since the higher-twist components  have a faster fall-off at small distances, the asymptotic results are not modified.

 Substituting  the pion wave function (\ref{eq:Phipi}) and using the integral representation for $V(Q^2, z)$
\begin{equation}
z Q K_1(z Q) = 2 Q^2 \int_0^\infty \frac{t J_0(z t)}{(t^2 + Q^2)^2} dt ,
\end{equation}
we find upon integration
\begin{equation}
F_{\pi \gamma}(Q^2) = \frac{\sqrt {2 P_{q \bar q}} ~ Q^2 }{\pi \kappa} \int_0^\infty \frac{t dt }{(t^2  + Q^2)^2} e^{- t^2 / 2 \kappa^2}.
\end{equation}
Changing variables as  $x = \frac{Q^2}{t^2 + Q^2}$ one obtains
\begin{equation}
F_{\pi \gamma}(Q^2) = \frac{P_{q \bar q}}{2 \pi^2 f_\pi} \int_0^1 dx \exp \left( - \frac{(1-x) P_{q \bar q} Q^2 }{4 \pi^2 f_\pi^2  x}\right).
\label{eq:FpiFCT2}
\end{equation}
Upon integration by parts, Eq. (\ref{eq:FpiFCT2})  can also be written as 
\begin{equation} \label{eq:TFFAdSQCD}
Q^2 F_{\pi \gamma}(Q^2) = \frac{4}{\sqrt{3}} \int_0^1 dx  \frac{\phi(x)}{1-x}  \left[1 - \exp \left( - \frac{(1-x) P_{q \bar q} Q^2 }{4 \pi^2 f_\pi^2  x}\right) \right] ,
\end{equation}
where $\phi(x) = \sqrt{3} f_\pi x(1-x)$ is the asymptotic QCD distribution amplitude with $f_\pi$ given by
(\ref{eq:fpiPhi}).

Remarkably, the pion transition form factor given by (\ref{eq:TFFAdSQCD}) 
for $P_{q \bar q} =1$
is identical to the 
results for the pion TFF obtained with the exponential light-front wave function model of 
Musatov and Radyushkin~\cite{Musatov:1997pu} consistent with the leading order  QCD result~\cite{Lepage:1980fj} for the TFF at the asymptotic limit,  
$Q^2 F_{\pi \gamma}(Q^2 \rightarrow \infty)=2 f_\pi$.~\footnote{A similar mapping can be done for the case when the two photons are virtual $\gamma^* \gamma^* \to \pi^0$.
In the case where at least one of the incoming photons has
large virtuality the transition form factor can be expressed analytically in a simple form. The result is 
$F_{\pi \gamma^*}(q^2, k^2) = - \frac{4}{\sqrt{3}} \int_0^1 dx  \frac{\phi(x)}{x q^2 + (1-x) k^2} $, with $\phi(x)$  the asymptotic DA. See Ref.~\cite{Grigoryan:2008up}.}
The leading-twist result  (\ref{eq:TFFAdSQCD}) does not include non-leading order  $\alpha_s$ corrections in the hard scattering amplitude nor gluon exchange in the evolution of
 the distribution amplitude, since the semiclassical correspondence implied in the gauge/gravity duality does not 
 contain quantum effects such as particle emission and absorption.~\footnote{The expression (\ref{eq:TFFAdSQCD}) is not appropriate to describe the time like region where the exponential factor in (\ref{eq:TFFAdSQCD}) grows exponentially. It is important to study the behavior of the pion TFF in other kinematical regions to describe, for example, the process $e^+ + e^- \to \gamma^* \to \pi^0 + \gamma$. This also would test the BaBar anomaly.}

The transition form factor at $Q^2=0$ can be obtained from Eq.~(\ref{eq:TFFAdSQCD}),
\begin{equation}
F_{\pi \gamma}(0) = \frac{1}{2  \pi^2 f_{\pi}} P_{q \bar q}.
\label{eq:pionTFFFCT2Q0}
\end{equation}
The form factor $F_{\pi \gamma}(0)$ is related to the decay width for the $\pi^0 \rightarrow \gamma \gamma$ decay,
\begin{equation}
\Gamma_{\pi^0 \rightarrow \gamma \gamma}=\frac{\alpha^2 \pi m_\pi^3}{4} F_{\pi \gamma}^2(0),
\end{equation}
where $\alpha=1/137$.
The form factor $F_{\pi \gamma}(0)$ is also well described by the Schwinger, Adler, Bell and Jackiw anomaly~\cite{Schwinger:1951nm} which gives
\begin{equation}
F_{\pi \gamma}^{\rm SABJ}(0) = \frac{1}{4 \pi^2 f_\pi},
\label{eq:JackiwAnomaly}
\end{equation}
 in agreement within a few percent of  the observed value obtained from the
the decay $\pi^0 \to \gamma \gamma$.

\begin{figure}[htbp]
\begin{center}
\includegraphics[width=9cm]{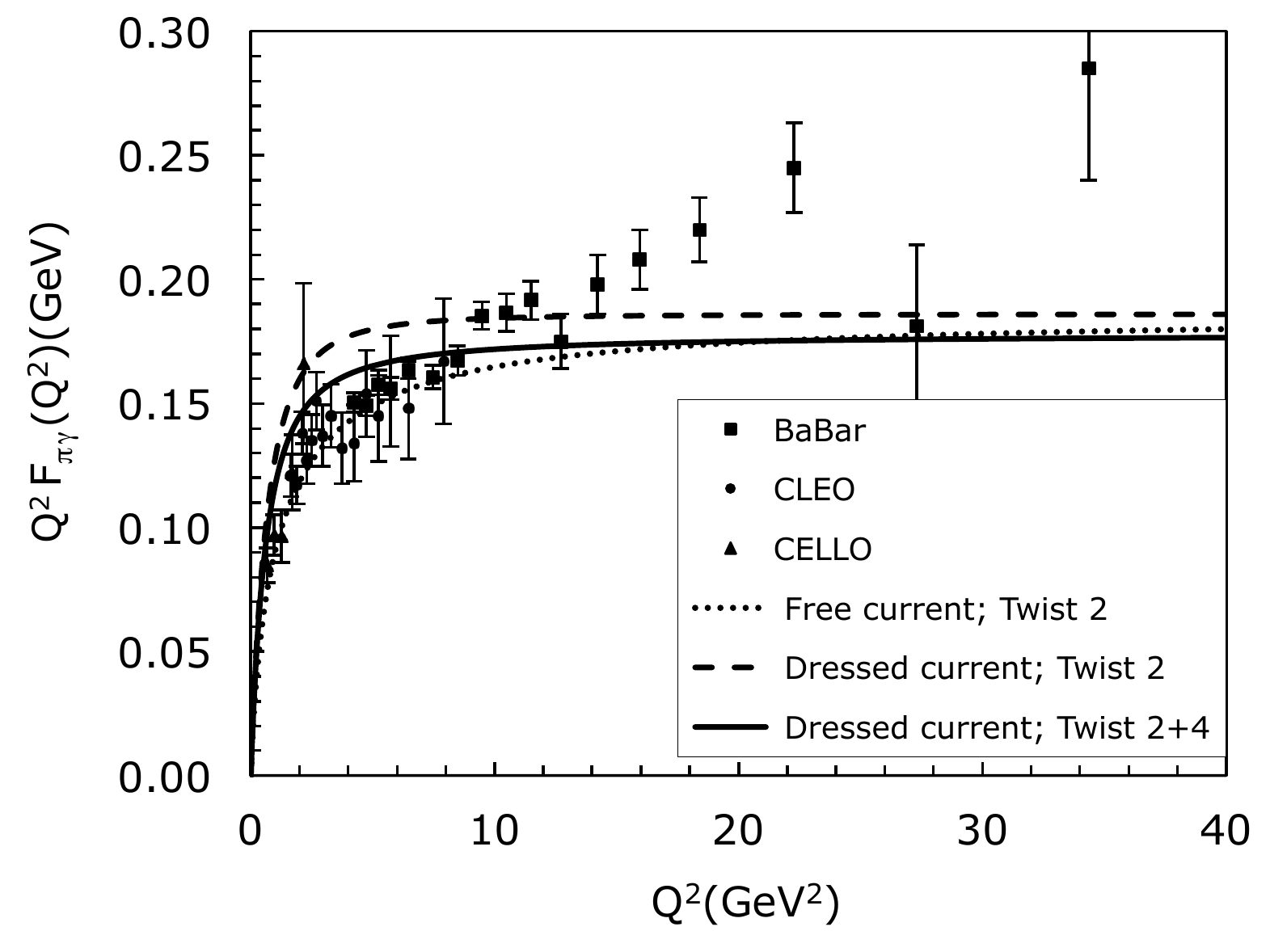}
\caption{The $\gamma \gamma^* \to \pi^0$ transition form factor shown as $Q^2 F_{\pi \gamma}(Q^2)$ as a function of $Q^2 = -q^2$.
The dotted curve is the asymptotic result predicted by the Chern-Simons form. 
The dashed and solid curves include the effects of using a confined EM current for twist-two and twist-two plus twist-four respectively. The data are from \cite{Aubert:2009mc, CELLO,  CLEO}. }
\label{Q2PiTFF}
\end{center}
\end{figure}

\begin{figure}[htbp]
\begin{center}
\includegraphics[width=9cm]{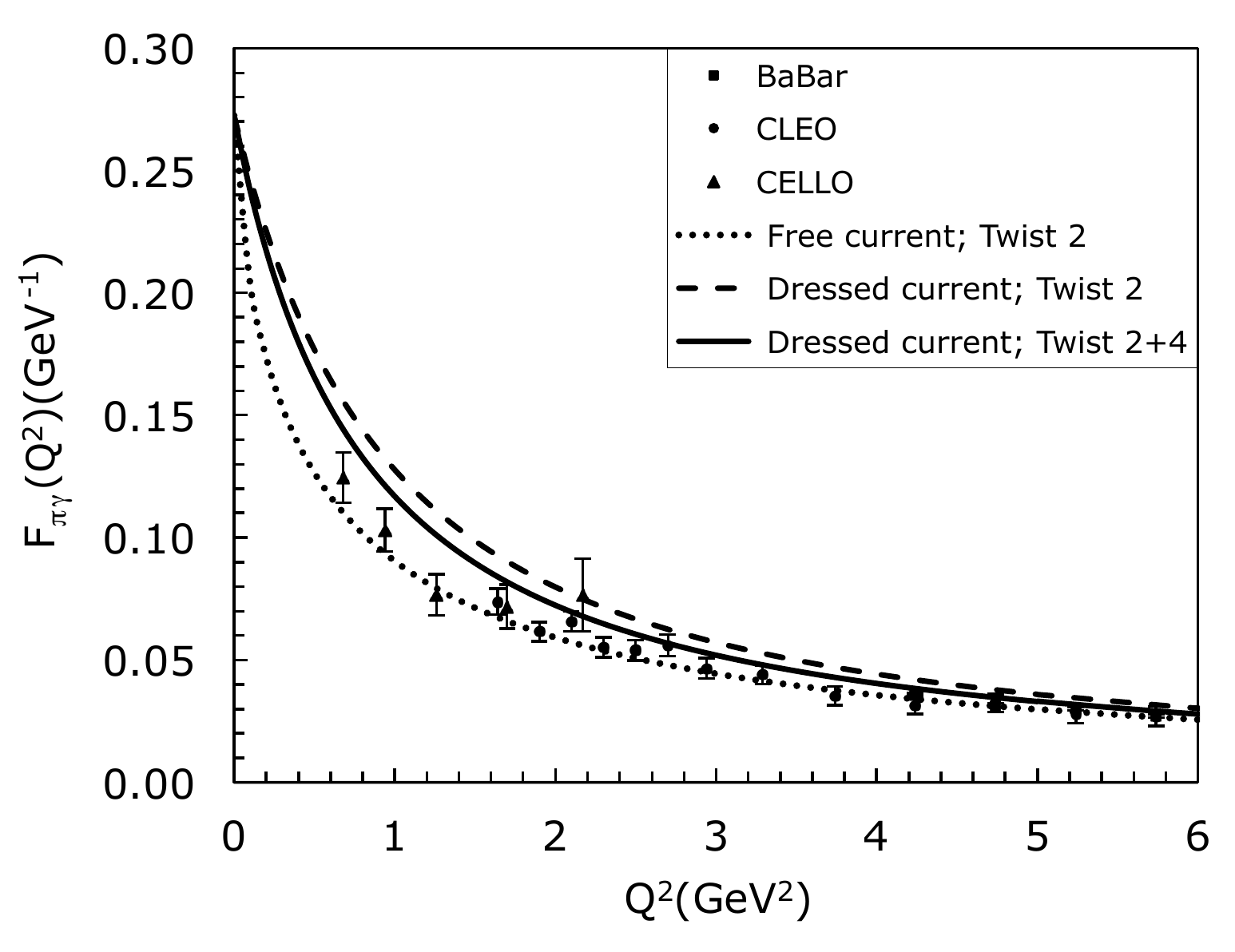}
\caption{Same as Fig.~\ref{Q2PiTFF} for $F_{\pi \gamma}(Q^2)$. }
\label{PiTFF}
\end{center}
\end{figure}

 Taking $P_{q \bar q}=0.5$ in (\ref{eq:pionTFFFCT2Q0}) one obtains a  result in agreement with (\ref{eq:JackiwAnomaly}).
 This suggests that the contribution from higher Fock states vanishes at $Q=0$ in this simple holographic confining model (see Section \ref{sec:HigherTwist} for further discussion).
 Thus (\ref{eq:TFFAdSQCD}) represents
 a description on the pion TFF
  which encompasses the low-energy non-perturbative  and the high-energy hard domains, but includes only
  the asymptotic DA of the $q \bar q$ component of the pion wave function at all scales.
The results from  (\ref{eq:TFFAdSQCD}) are shown as dotted curves in Figs.  \ref{Q2PiTFF} and  \ref{PiTFF} for $Q^2 F_{\pi\gamma}(Q^2)$ and $F_{\pi\gamma}(Q^2)$ respectively.
The calculations agree reasonably well with the experimental data at low- and medium-$Q^2$ regions ($Q^2<10$ GeV$^2$) , but disagree with BaBar's large $Q^2$ data.

\subsection{Transition Form Factor with the Dressed AdS/QCD Current}
\label{sec:DressedCurrent}

The simple valence $q \bar q$ model discussed above should be modified at small $Q^2$ by introducing the  dressed current  which corresponds effectively to a superposition of Fock states.
In the case of soft-wall potential,~\cite{Karch:2006pv}
the EM bulk-to-boundary propagator is~\cite{Brodsky:2007hb, Grigoryan:2007my}
\begin{equation} \label{eq:Vkappa}
V(Q^2,z) = \Gamma\left(1 + \frac{Q^2}{4 \kappa^2}\right) U\left(\frac{Q^2}{4 \kappa^2}, 0, \kappa^2 z^2\right),
\end{equation}
where $U(a,b,c)$ is the Tricomi confluent hypergeometric function.  The modified current $V(Q^2,z)$, Eq. (\ref{eq:Vkappa}),  has the same boundary conditions as the free current (\ref{eq:intJ}),
and reduces to (\ref{eq:intJ}) in the  limit $Q^2 \to \infty$.
Eq.~(\ref{eq:Vkappa}) can be conveniently written in terms of the integral representation~\cite{Grigoryan:2007my}
\begin{equation}  \label{eq:Vx}
V(Q^2,z) = \kappa^2 z^2 \int_0^1 \! \frac{dx}{(1-x)^2} \, x^{\frac{Q^2}{4 \kappa^2}} 
e^{-\kappa^2 z^2 x/(1-x)}.
\end{equation}

Inserting the valence pion wave function (\ref{eq:Phipi})  and the confined EM current  (\ref{eq:Vx}) in 
the amplitude (\ref{eq:TFFAdS2})
one finds 
\begin{equation}  \label{eq:DressedFpiT2}
F_{\pi \gamma}(Q^2) = \frac{P_{q \bar q} }{\pi^2 f_\pi} \int_0^1  \frac{d x }{(1+x)^2} \, x^{Q^2 P_{q \bar q}/(8 \pi^2 f_\pi^2)}.
\end{equation}
Eq.~(\ref{eq:DressedFpiT2}) gives the same value for $F_{\pi \gamma}(0)$ as (\ref{eq:pionTFFFCT2Q0}) which was  obtained with the free current.
Thus the anomaly result $F_{\pi \gamma}(0)=1/(4 \pi^2 f_\pi)$ is reproduced if $P_{q \bar q}=0.5$ is also taken in (\ref{eq:DressedFpiT2}).
 Upon integration by parts, Eq.~(\ref{eq:DressedFpiT2}) can also be written as
\begin{equation}
Q^2 F_{\pi \gamma}(Q^2) = 8 f_\pi  \int_0^1  d x \frac{1- x }{(1+x)^3} \, \left( 1- x^{Q^2 P_{q \bar q}/(8 \pi^2 f_\pi^2)} \right).
\label{eq:DressedFpiT2Asy}
\end{equation}
Noticing that the second term in Eq.~(\ref{eq:DressedFpiT2Asy}) vanishes at the limit $Q^2 \rightarrow \infty$,
one recovers Brodsky-Lepage's asymptotic prediction for the pion TFF:  $Q^2 F_{\pi \gamma}(Q^2 \rightarrow \infty)=2 f_\pi$.~\cite{Lepage:1980fj}

The results calculated with (\ref{eq:DressedFpiT2}) for $P_{q \bar q}=0.5$ are shown as dashed curves in Figs.  \ref{Q2PiTFF} and  \ref{PiTFF}.
One can see that the calculations with the dressed current are larger as compared with the results computed with the free current and the experimental data at low- and medium-$Q^2$
regions ($Q^2 <10$ GeV$^2$). The new results again disagree with BaBar's data at large $Q^2$.

\subsection{Higher-Twist Components to the Transition Form Factor}
\label{sec:HigherTwist}

In a previous light-front QCD analysis of the pion TFF \cite{Brodsky:1980vj} it was argued that the valence Fock state $| q \bar q \rangle$ provides only half of the contribution to
the pion TFF at $Q^2=0$, while the other half comes from diagrams where the virtual photon couples inside the pion
(strong interactions occur between the two photon interactions).
This leads to a surprisingly small value for the valence Fock state probability $P_{q \bar q}=0.25$.
More importantly, this raises the question on the role played by the higher Fock components of the pion LFWF,
\begin{equation}
\vert \pi \rangle =  \psi_2 \vert \bar q q \rangle + \psi_3 |q \bar q g \rangle +  \psi_4 | q \bar q q \bar q \rangle + \cdots,
\label{eq:Fockexp}
\end{equation}
in the calculations for the pion TFF.

The contributions to the transition form factor from these higher Fock states are suppressed,  compared with the valence Fock state, by the factor $1/(Q^2)^n$
for $n$ extra  $q \bar q$ pairs in the higher Fock state, since one needs to evaluate an off-diagonal matrix element between the real photon
and the multi-quark Fock state.~\cite{Lepage:1980fj}
We note that in the case of the elastic form factor the power suppression is $1/ (Q^2)^{2n}$ for $n$ extra $q \bar q$ pairs in the higher Fock state.
These higher Fock state contributions are negligible at high $Q^2$.
On the other hand, it has  long been argued  that the higher Fock state
contributions are necessary to explain the experimental data at the medium $Q^2$ region for exclusive
processes.~ \cite{CaoHM96,CaoCHM97}
The contributions from the twist-3 parts of the two-parton pion distribution amplitude to the pion elastic form factors were evaluated
in Ref. \cite{CaoDH99}.  The three-parton contributions to the pion elastic  form factor were studied in Ref. \cite{ChenL11}.
The contributions from diagrams where the virtual photon couples inside the pion to the pion transition form factor were estimated using light-front wavefunctions in Ref. \cite{HuangW07,WuH10}.
The higher twist (twist-4 and twist-6) contributions to the pion transition form factor~\cite{Gorsky87} were evaluated using the method of light-cone sum rules in Refs.~\cite{SAgaevBOP11,BakulevMPS11},
 but opposite claims were made on whether the BaBar\ data could be accommodated by including these higher twist contributions.

It is also not very clear how the higher Fock states contribute to decay processes, such as $\pi^0 \rightarrow \gamma \gamma$, \cite{AlkoferR92}
due to the long-distance non-perturbative nature of decay processes.
Second order radiative corrections to the triangle anomaly do not change the anomaly results as they contain one internal photon line and two vertices on the triangle loop.
Upon regulation no new anomaly contribution occurs.
In fact, the result is expected to be valid at all orders in perturbation theory.~\cite{Adler:1969er, Zee:1972zt}
It is thus generally argued that in the chiral limit of QCD ({\it i.e.}, $m_q \rightarrow 0$), one needs only the $q \bar q$ component to explain the anomaly, but as shown below,
the higher Fock state components
 can also contribute to the decay process $\pi^0 \to \gamma \gamma$ in the chiral limit.

As discussed in the last two sections, matching the AdS/QCD results computed with the free and dressed currents for the TFF at $Q^2=0$ with the anomaly result
requires a probability $P_{q \bar q}=0.5$.~\footnote{The asymptotic normalization of the pion form factor 
$Q^2 F_\pi(Q^2 \to \infty)$ is dependent on the valence probability  $P_{q \bar q}$ and thus changes effectively the
mass scale by a factor $\sqrt{2}$.~\cite{Brodsky:2011xx}}
Thus it is important to investigate the contributions from the higher Fock states.
In AdS/QCD there are no dynamic gluons and confinement is realized via an effective instantaneous interaction
in light-front time, analogous to the instantaneous gluon exchange.~\cite{Brodsky:1997de}
The effective confining  potential also creates quark-antiquark pairs from the amplitude $q \to q \bar q q$.
Thus  in AdS/QCD higher Fock states can have any number of extra $q \bar q$ pairs.
These higher Fock states lead to higher-twist contributions to the pion transition form factor.

\begin{figure}[h!]
\begin{center}
\includegraphics[width=12cm]{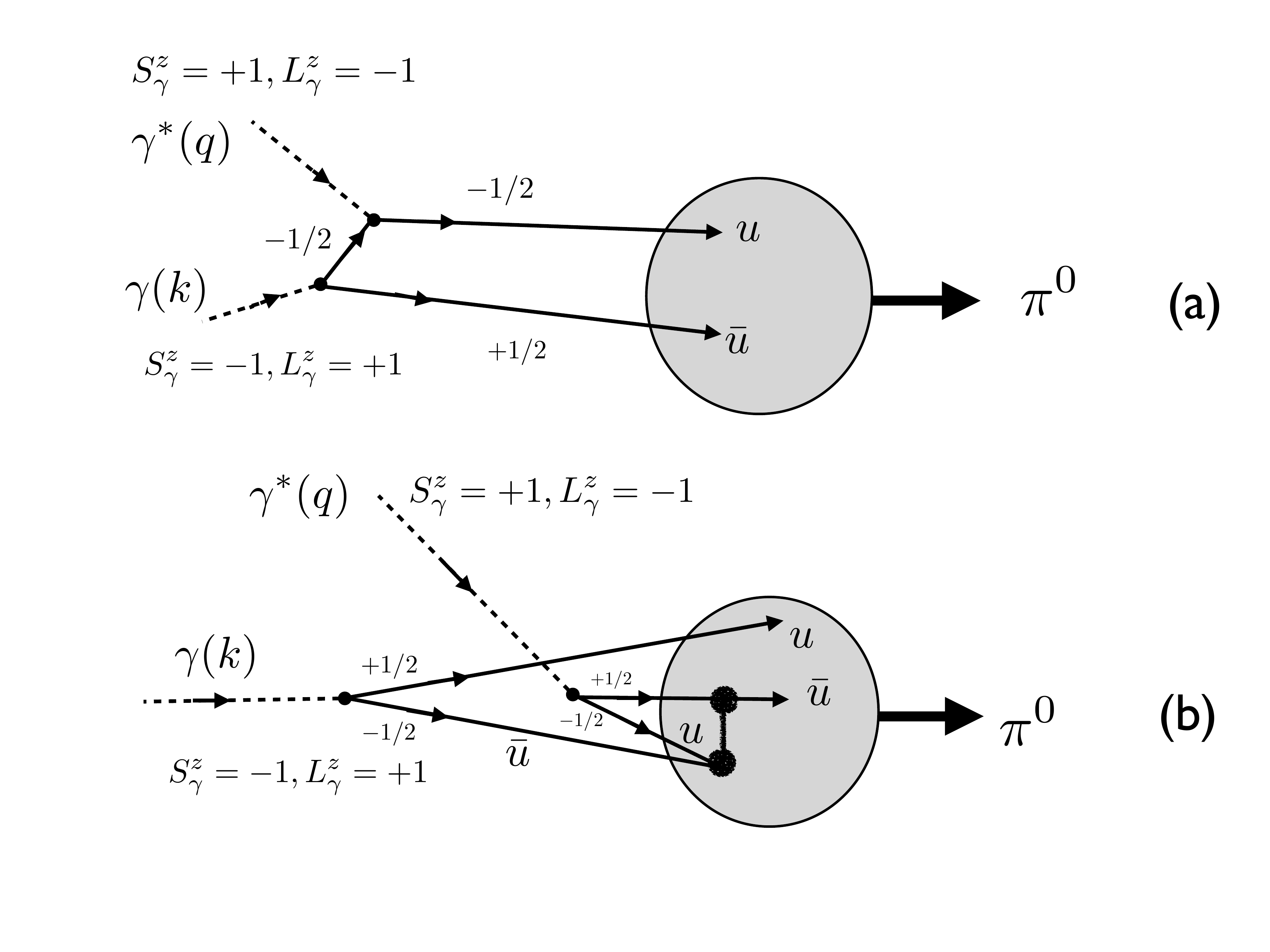}
\caption{Leading-twist contribution (a) and twist-four contribution (b) to the process $\gamma \gamma^* \to \pi^0$.}
\label{TFF}
\end{center}
\end{figure}

To illustrate this observation consider the two diagrams in Fig.~\ref{TFF}.
In the leading process,  Fig.~\ref{TFF} (a), where  both photons couple to the same quark, the valence
$|q \bar q \rangle$ state has  $J^z = S^z = L^z =0$,
\begin{equation} \label{eq:2q}
\vert q \bar q \rangle = \frac{1}{\sqrt 2} \Big( \Big\vert + \half, - \half \Big\rangle -  \Big\vert - \half, + \half \Big\rangle
\Big).
\end{equation}
Eq.~(\ref{eq:2q}) represents a $J^{PC} = 0^{-+}$ state with the quantum numbers of the
conventional $\pi$ meson
axial vector interpolating operator $\mathcal{O} = \bar \psi \gamma^+ \gamma^5 \psi$.

In the process involving the four quark state $\vert q \bar q q \bar q\rangle$ of the pion, Fig.~\ref{TFF} (b), where each photon
couples directly to a $q \bar q$ pair, the four quark state also satisfies $J^z = S^z = L^z =0$ and is represented by
\begin{multline} \label{eq:4q}
\vert q \bar q q \bar q \rangle = \frac{1}{ 2} \Big(
     \Big\vert + \half, - \half, + \half, - \half \Big\rangle
+  \Big\vert + \half, - \half, - \half, + \half \Big\rangle \\
-   \Big\vert - \half, + \half, + \half, - \half \Big\rangle
-   \Big\vert - \half, + \half, - \half,  + \half \Big\rangle \Big).
\end{multline}
The four quark state in Eq.~(\ref{eq:4q}) has also quantum numbers $J^{PC} = 0^{-+}$ corresponding to  the quantum numbers of the local interpolating operators
$\mathcal{O} = \bar \psi \gamma^+ \gamma^5 \psi \psi \bar \psi$ where
the  scalar interpolating operator $\bar \psi \psi$ has quantum numbers $J^{PC} = 0^{++}$.

We note that for the Compton scattering $\gamma H \to \gamma H$ process, similar higher-twist contributions, as illustrated in Fig.~\ref{TFF} (b),
are proportional to $\sum_{e_i \ne e_j} e_i e_j$ and are necessary to derive the low energy amplitude for Compton scattering
which is proportional to the total charge squared $e_H^2 = (e_i + e_j)^2$ of the target. \cite{Brodsky:1968ea}

Both processes illustrated in Fig~(\ref{TFF}) make contributions to the two photon process $\gamma^* \gamma \to \pi^0$.
Time reversal invariance means that the four quark state $\vert q \bar q q \bar q\rangle$ should also contribute to the decay process $\pi^0 \to \gamma \gamma$.
In a semiclassical model without dynamic gluons, Fig.~\ref{TFF} (b) represents the only higher twist term which contribute to the $\gamma^* \gamma \to \pi^0$ process.
The twist-four contribution vanishes at large $Q^2$ compared to the leading-twist  contribution, thus maintaining the asymptotic predictions while only modifying
the large distance behavior of the wave function.

To investigate the contributions from the higher Fock states in the pion LFWF, we write the twist-two and twist-four hadronic AdS components from (\ref{eq:Phitau})
\begin{eqnarray}  \label{eq:t2}
\Phi_\pi^{\tau = 2}(z) &=& \frac{\sqrt 2 \kappa z^2 }{\sqrt{1 + \alpha^2}}  e^{- \kappa^2 z^2/2}, \\ \label{eq:t4}
\Phi_\pi^{\tau = 4}(z) &=&  \frac{\alpha \kappa^3 z^4 }{\sqrt{1 + \alpha^2}}  e^{- \kappa^2 z^2/2},
\end{eqnarray}
with normalization
\begin{equation}
\int_0^\infty \frac{dz}{z^3} \left[\vert \Phi_\pi^{\tau = 2}(z) \vert^2 + \vert \Phi_\pi^{\tau = 4}(z) \vert^2 \right] = 1,
\label{eq:normT24}
\end{equation}
and probabilities $P_{q \bar q}=1/(1+ \vert \alpha \vert^2)$ and $P_{q \bar q q \bar q}=\alpha^2/(1+ \vert \alpha \vert^2)$.
The pion decay constant
follows from the short distance asymptotic behavior of the leading contribution
and is given by
\begin{equation}  \label{eq:fpialpha}
f_\pi = \frac{1}{\sqrt{1 + \alpha^2}} \frac{\kappa}{ \sqrt 2 \pi}.
\end{equation}

Using (\ref{eq:t2}) and (\ref{eq:t4}) together with (\ref{eq:Vx}) in equation (\ref{eq:TFFAdS2}) we find the total contribution from twist-two and twist-four components for the
dressed current,
\begin{equation}
F_{\pi \gamma}(Q^2) =  \frac{1 }{\pi^2 f_\pi}  \frac{1}{(1 + \alpha^2)^{3/2}} \int_0^1  \frac{d x }{(1+x)^2}
x^{Q^2 / [8 \pi^2 f_\pi^2 (1+\alpha^2)]}
\left[1 + \frac{4 \alpha}{\sqrt 2} \frac{1-x}{1+x} \right].
\label{eq:DressedFpiT2+4}
\end{equation}
The transition from factor at $Q^2=0$ is given by
\begin{equation}
F_{\pi \gamma}(0)=\frac{1 }{2 \pi^2 f_\pi}  \frac{1+\sqrt{2} \alpha }{(1 + \alpha^2)^{3/2}}.
\label{eq:FpiTwist2+4Q0}
\end{equation}

Imposing the anomaly result (\ref{eq:JackiwAnomaly}) on (\ref{eq:FpiTwist2+4Q0}) we find two possible real solutions for $\alpha$: $\alpha_1=-0.304$ and $\alpha_2=1.568$.
The larger value $\alpha_2=1.568$ yields $P_{q \bar q}=0.29$, $P_{q \bar q q \bar q}=0.71$, and $\kappa=1.43$ GeV. The  resulting value of $\kappa$ is about 4 times larger than the value
obtained from the AdS/QCD analysis of the hadron spectrum and the pion elastic form factor,
and thereby should be discarded.
The other solution $\alpha_1=-0.304$ gives $P_{q \bar q}=0.915$, $P_{q \bar q q \bar q}=0.085$, and $\kappa=0.432$ GeV -- results that
are similar to that found from an analysis of the space and time-like behavior of the pion form factor using LF holographic methods,~\footnote{If we impose the condition that the twist 4 contribution at $Q^2=0$ is exactly half the value of the twist 2 contribution one obtains $\alpha = - \frac{1}{2 \sqrt 2}$, which is very close to the value of $\alpha$ which follows by imposing the
  triangle anomaly constraint. In this case the pion TFF has a very simple form
  $F_{\pi \gamma}(Q^2) = \frac{8 }{3 \pi \kappa} \int_0^1  \frac{d x}  {(1+x)^3} \, x^{Q^2 / 4 \kappa^2 + 1}$.}
including higher Fock components in the pion wave function.~\cite{deTeramond:2010ez}
Semiclassical holographic methods, where dynamical gluons are not presented, are thus compatible with a large probability for the valence state of the order of 90\%.
On the other hand, QCD analyses including multiple gluons on the pion wave function favor a small probability (25\%) for the valence
state.~\cite{Brodsky:1980vj} Both cases (and examples in between) are examined in Ref.~\cite{Cao:2011}.

The results for the transition form factor are shown as solid curves in Figs.  \ref{Q2PiTFF} and  \ref{PiTFF}.
The agreements with the experimental data at low- and medium-$Q^2$ regions ($Q^2 < 10$ GeV$^2$) are greatly improved compared with the results obtained with only
twist-two component computed with the dressed current.
However, the rapid growth of the pion-photon transition form factor exhibited by the BaBar\  data at high $Q^2$ still cannot be reproduced.
So we arrive at a similar conclusion as we did in a QCD analysis of the pion TFF in Ref.~\cite{Cao:2011}:
it is difficult to explain the rapid growth of the form factor exhibited by the BaBar\  data at high $Q^2$ within the current framework of QCD.

\subsection{Transition Form Factors for the $\eta$ and $\eta^\prime$ Mesons}
\label{sec:eta}

The  $\eta$ and $\eta^\prime$ mesons result from the mixing of
the neutral states $\eta_8$ and $\eta_1$ of the SU(3)$_F$ quark model. The transition form factors for the latter have the same expression as  the pion transition form factor,
except an overall multiplying factor $c_P=1, \, \frac{1}{\sqrt{3}}$, and $\frac{2\sqrt{2}}{\sqrt{3}}$ for the $\pi^0$, $\eta_8$ and $\eta_1$, respectively.
By multiplying  equations (\ref{eq:TFFAdSQCD}), (\ref{eq:DressedFpiT2}) and (\ref{eq:DressedFpiT2+4})
by the appropriate factor  $c_P$, one obtains the corresponding expressions for the transition form factors for the $\eta_8$ and $\eta_1$.

\begin{figure}[htbp][h]
\begin{center}
\includegraphics[width=9cm]{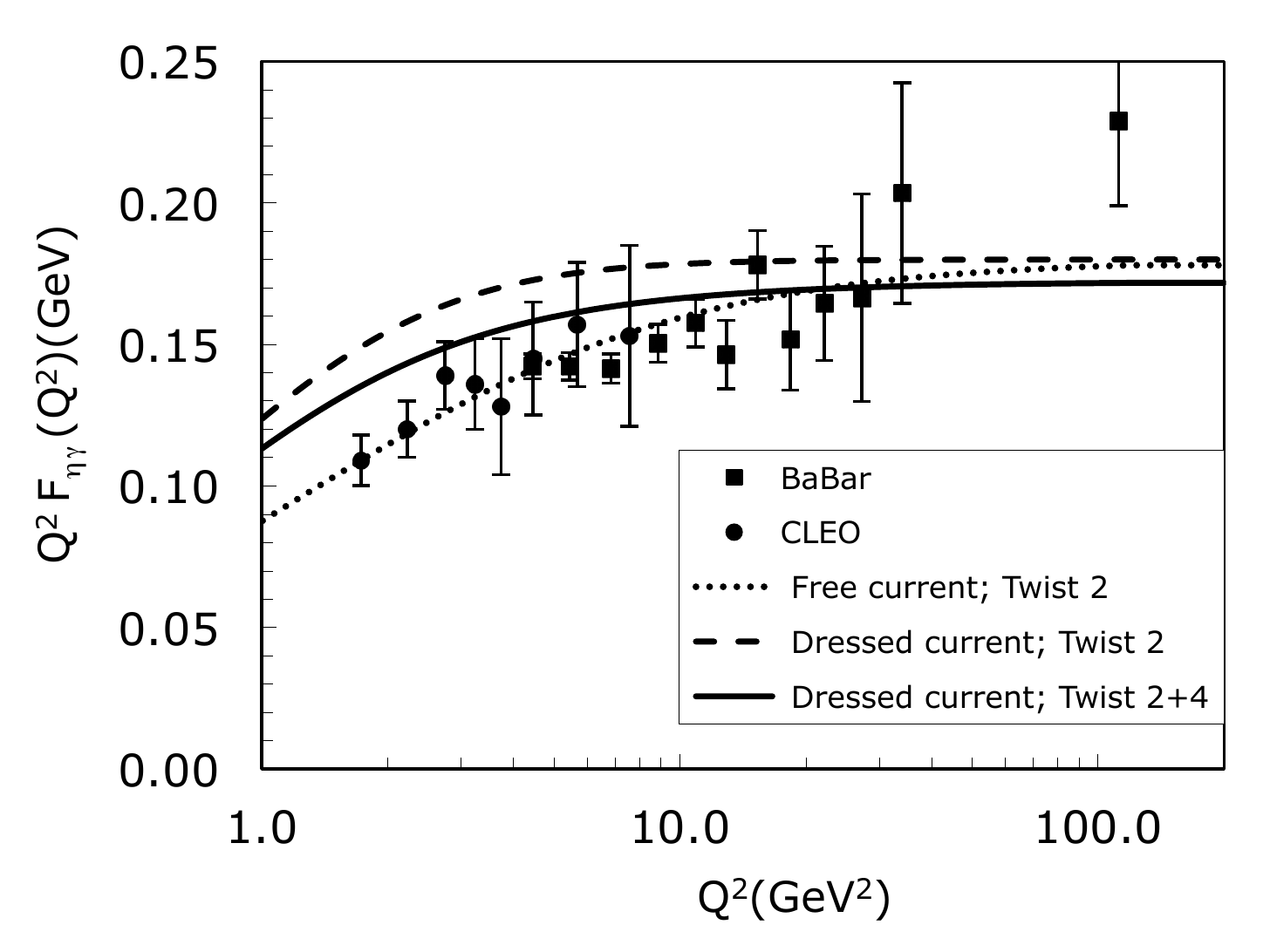}
\caption{The $\gamma \gamma^* \rightarrow \eta$ transition form factor shown as $Q^2 F_{\eta \gamma}(Q^2)$ as a function of $Q^2 = -q^2$.  The dotted curve is the asymptotic result.
The dashed and solid curves include the effects of using a confined EM current for twist two and twist two plus twist four respectively. The data are from \cite{Aubert:2009mc, CELLO,  CLEO}. }\label{fig:EtaTFF_Q2Feta}
\end{center}
\end{figure}

\begin{figure}[htbp]
\begin{center}
\includegraphics[width=9cm]{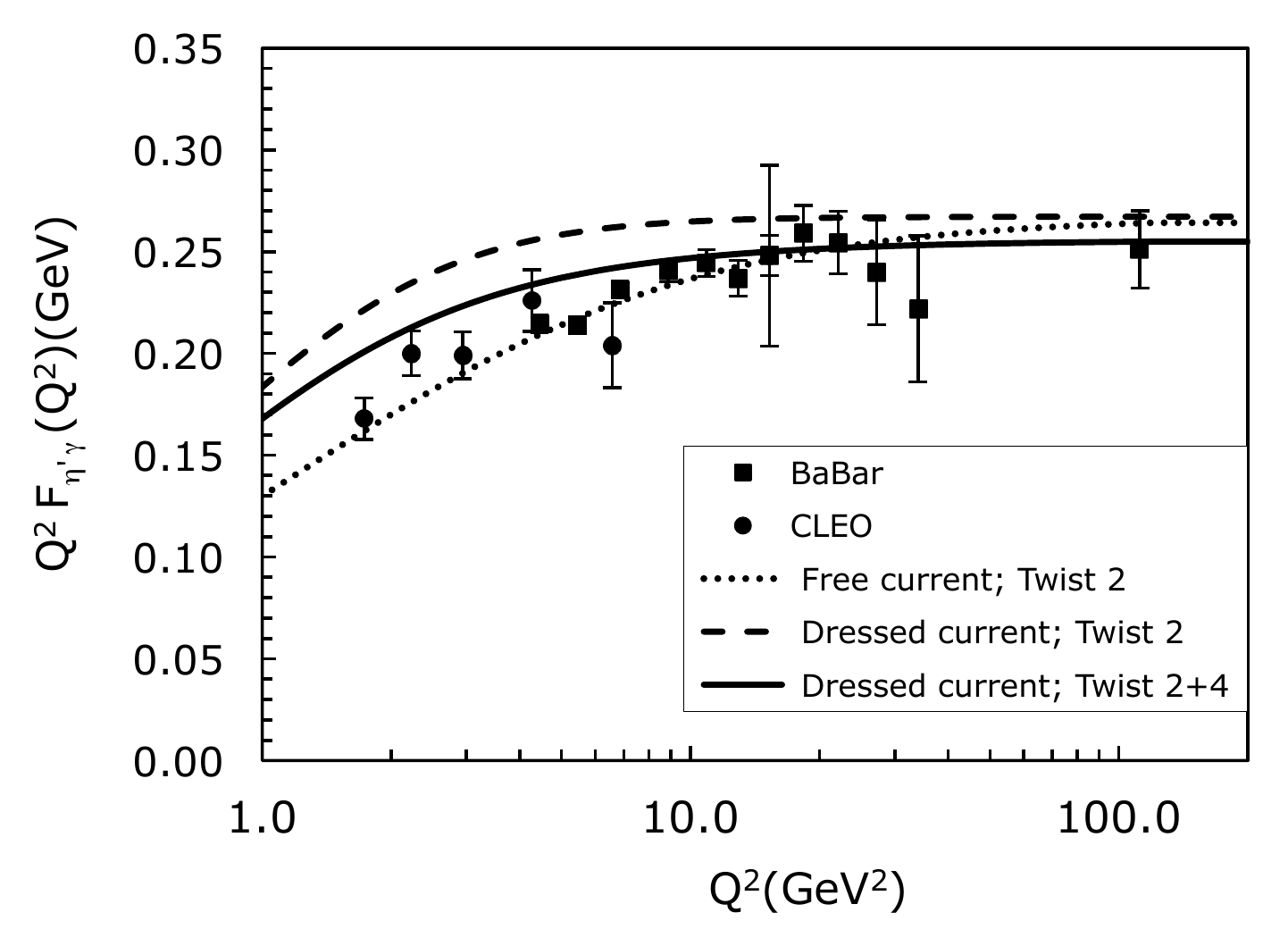}
\caption{Same as Fig. \ref{fig:EtaTFF_Q2Feta} for the $\gamma \gamma^* \rightarrow \eta^\prime$ transition form factor
shown as   $Q^2 F_{\eta^\prime \gamma}(Q^2)$.}
\label{fig:EtaPTFF_Q2FetaP}
\end{center}
\end{figure}

\begin{figure}[htbp]
\begin{center}
\includegraphics[width=9cm]{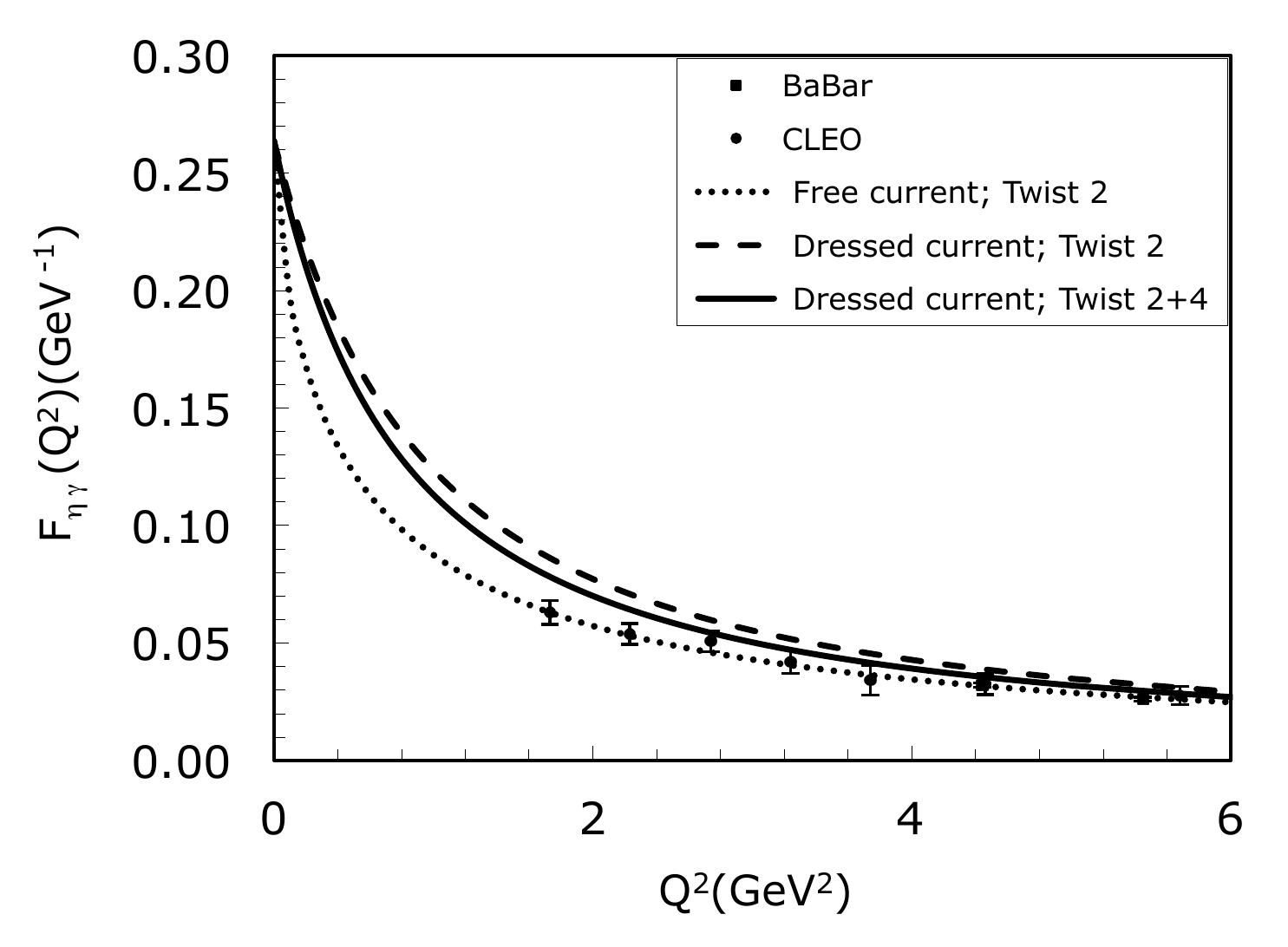}
\caption{Same as Fig. \ref{fig:EtaTFF_Q2Feta} for the $\gamma \gamma^* \rightarrow \eta$ transition form factor
shown as   $F_{\eta \gamma}(Q^2)$.}
\label{fig:EtaTFF_Feta}
\end{center}
\end{figure}

\begin{figure}[htbp]
\begin{center}
\includegraphics[width=9cm]{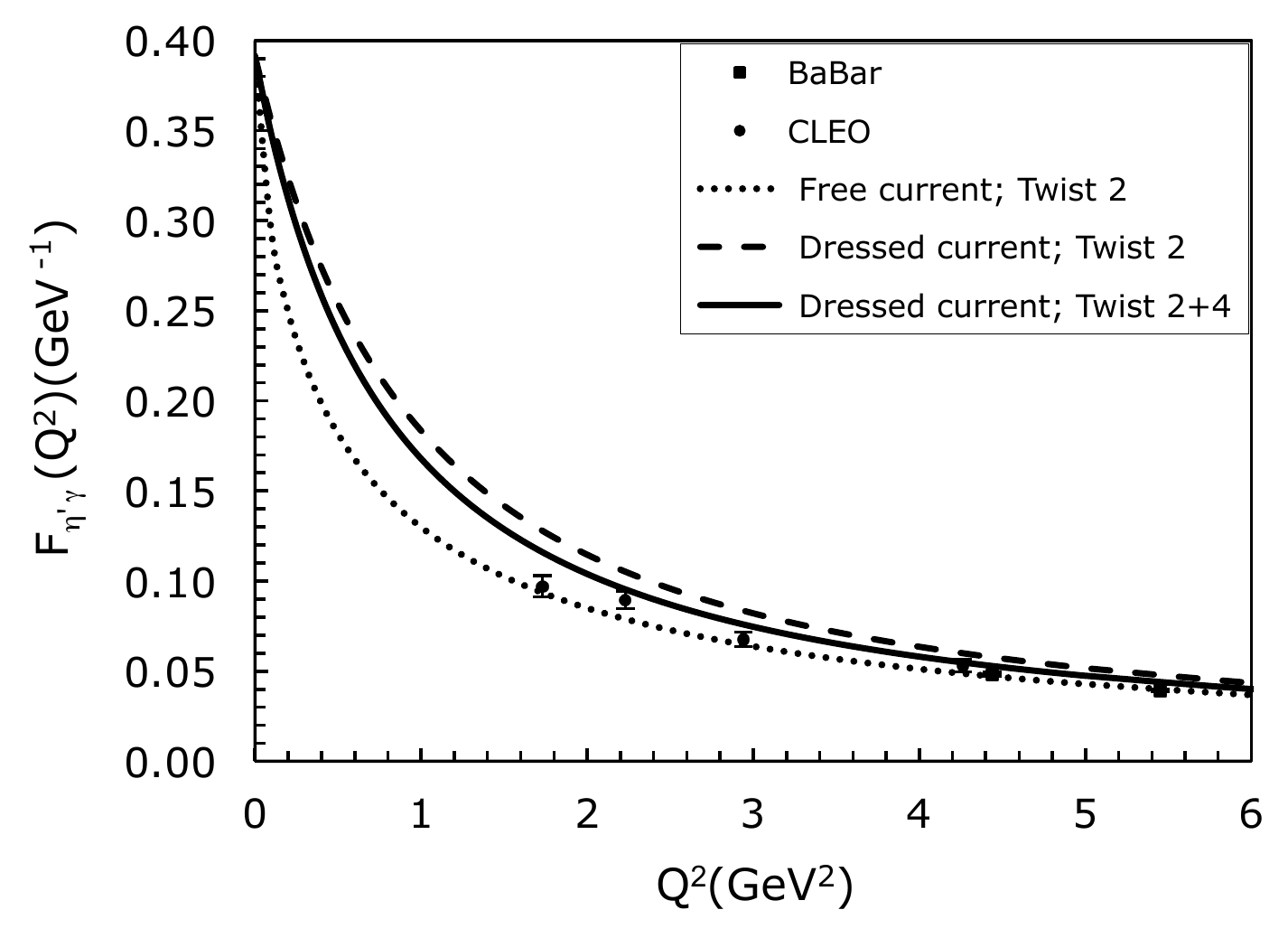}
\caption{Same as Fig. \ref{fig:EtaTFF_Q2Feta} for the $\gamma \gamma^* \rightarrow \eta^\prime$ transition form factor
shown as   $F_{\eta^\prime \gamma}(Q^2)$.}
\label{fig:EtaPTFF_FetaP}
\end{center}
\end{figure}

The transition form factors for the physical states  $\eta$ and $\eta^\prime$ are a superposition of the transition form factors for the  $\eta_8$ and $\eta_1$
\begin{eqnarray}
\left(
	\begin{array}{c}
	F_{\eta \gamma} \\
	F_{\eta^\prime \gamma}
	\end{array}
\right)
=\left(
	\begin{array}{cc}
	{\rm cos} \, \theta  & -{\rm sin} \, \theta \\
	{\rm sin} \, \theta & {\rm cos} \, \theta
	\end{array}
\right)
\left(
	\begin{array}{c}
	F_{\eta_8 \gamma} \\
	F_{\eta_1 \gamma}
	\end{array}
\right),
\end{eqnarray}
where $\theta$ is the mixing angle for which we adopt $\theta=-14.5^o\pm 2^o$.~\cite{Cao:1999fs}
The results for the $\eta$ and $\eta^\prime$ transitions form factors are shown in
Figs. \ref{fig:EtaTFF_Q2Feta} and \ref{fig:EtaPTFF_Q2FetaP} for $Q^2 F_{M\gamma}(Q^2)$, and Figs. \ref{fig:EtaTFF_Feta} and \ref{fig:EtaPTFF_FetaP} for $F_{M\gamma}(Q^2)$.
The calculations agree very well with available experimental data over a large range of $Q^2$.
We note that other mixing schemes were proposed in studying the mixing behavior of the decay constants and states of the $\eta$ and $\eta^\prime$ mesons. \cite{Leutwyler98,FeldmannK98,FeldmannKS98}
Since the transition from factors are the primary interest in this study it is appropriate to use the conventional single-angle mixing scheme
for the states. Furthermore, the predictions for the $\eta$ and $\eta^\prime$ transition form factors remain largely unchanged if other mixing schemes are used in the calculation.

\section{Light-Front Bound-State Hamiltonian Equation of Motion}

A key step in the analysis of an atomic system such as positronium
is the introduction of the spherical coordinates $r, \theta, \phi$
which  separates the dynamics of Coulomb binding from the
kinematical effects of the quantized orbital angular momentum $L$.
The essential dynamics of the atom is specified by the radial
Schr\"odinger equation whose eigensolutions $\psi_{n,L}(r)$
determine the bound-state wavefunction and eigenspectrum. In our recent
work, we have shown that there is an analogous invariant
light-front coordinate $\zeta$ which allows one to separate the
essential dynamics of quark and gluon binding from the kinematical
physics of constituent spin and internal orbital angular momentum.
The result is a single-variable LF Schr\"odinger equation for QCD
which determines the eigenspectrum and the light-front wavefunctions
of hadrons for general spin and orbital angular momentum.~\cite{deTeramond:2008ht}  If one further chooses  the constituent rest frame (CRF)~\cite{Danielewicz:1978mk,Karmanov:1979if,Glazek:1983ba}  where $\sum^n_{i=1} \mbf{k}_i \! = \! 0$, then the kinetic energy in the LFWF displays the usual 3-dimensional rotational invariance. Note that if the binding energy is nonzero, $P^z \ne 0,$ in this frame.

One can also derive light-front holography using a first semiclassical approximation  to transform the fixed
light-front time bound-state Hamiltonian equation of motion in QCD
\begin{equation} \label{LFH}
H_{LF} \vert  \psi(P) \rangle =  M_{H}^2 \vert  \psi(P) \rangle,
\end{equation}
with  $H_{LF} \equiv P_\mu P^\mu  =  P^- P^+ -  \mbf{P}_\perp^2$,
to  a corresponding wave equation in AdS
space.~\cite{deTeramond:2008ht} To this end we
 compute the invariant hadronic mass $M^2$ from the hadronic matrix element
\begin{equation}
\langle \psi_H(P') \vert H_{LF} \vert \psi_H(P) \rangle  =
M_H^2  \langle \psi_H(P' ) \vert\psi_H(P) \rangle,
\end{equation}
expanding the initial and final hadronic states in terms of its Fock components. We use the
frame $P = \big(P^+, M^2/P^+, \vec{0}_\perp \big)$ where $H_{LF} =  P^+ P^-$.
We find
\begin{equation}
 M_H^2  =  \sum_n  \prod_{j=1}^{n-1} \int d x_j \, d^2 \mbf{b}_{\perp j} \,
\psi_{n/H}^*(x_j, \mbf{b}_{\perp j})
  \sum_q   \left(\frac{ \mbf{- \nabla}_{ \mbf{b}_{\perp q}}^2  \! + m_q^2 }{x_q} \right)
 \psi_{n/H}(x_j, \mbf{b}_{\perp j})
  + {\rm (interactions)} , \label{eq:Mba}
 \end{equation}
plus similar terms for antiquarks and gluons ($m_g = 0)$.

 Each constituent of the light-front wavefunction  $\psi_{n/H}(x_i, \mbf{k}_{\perp i}, \lambda_i)$  of a hadron is on its respective mass shell
 $k^2_i= k^+_i k^-_i - \mbf{k}^2_\perp = m^2_i$, $i = 1, 2 \cdots n,$
and thus $k^-= {{\mbf k}^2_\perp+  m^2_i\over x_i P^+}$.
However,  the light-front wavefunction represents a state which is off the light-front energy shell: $P^-  - \sum_i^n k^-_n < 0$, for a stable hadron.  Scaling out $P^+ = \sum^n_i k^+_i$, the
off-shellness of the $n$-parton LFWF is thus $M^2_H -  M^2_n$, where
the invariant mass  of the constituents $M_n$ is
\begin{equation}
 M_n^2  = \Big( \sum_{i=1}^n k_i^\mu\Big)^2 = \sum_i \frac{\mbf{k}_{\perp i}^2 +  m_i^2}{x_i}.
 \end{equation}

The action principle selects the configuration which minimizes the time-integral of the Lagrangian $L= T-V$, thus minimizing the kinetic energy $T$  and maximizing the attractive forces of the potential $V$. Thus in a fixed potential, the light-front wavefunction  peaks at the minimum value of the invariant mass of the constituents; i.e.,  at  the minimum off-shellness $M^2_H -  M^2_n$.   In the case of massive constituents,  the minimum LF off-shellness occurs when all of the constituents have equal rapidity: $x_i \simeq {m^2_{\perp i}\over \sum^n_j m^2_{\perp j}}, $ where $m_{\perp i} = \sqrt {k^2_{\perp i} + m^2_i}.$ This is the central principle underlying the intrinsic heavy sea-quark distributions of hadrons.  The functional dependence  for a given Fock state is given in terms of the invariant mass,  the measure of the off-energy shell of the bound state.

If we want to simplify further the description of the multiple parton system and reduce its dynamics to a single variable problem, we must take the limit of quark masses to zero.
Indeed, the underlying classical QCD Lagrangian with massless quarks is scale and conformal invariant,~\cite{Parisi:1972zy} and consequently only in this limit it is possible to map the equations of motion and transition matrix elements to their correspondent conformal AdS  expressions.

 To simplify the discussion we will consider a two-parton hadronic bound state.  In the limit of zero quark masses
$m_q \to 0$
\begin{equation}  \label{eq:Mb}
M^2  =  \int_0^1 \! \frac{d x}{x(1-x)} \int  \! d^2 \mbf{b}_\perp  \,
  \psi^*(x, \mbf{b}_\perp)
  \left( - \mbf{\nabla}_{ {\mbf{b}}_{\perp}}^2\right)
  \psi(x, \mbf{b}_\perp) +   {\rm (interactions)}.
 \end{equation}
For $n=2$, $M_{n=2}^2 = \frac{\mbf{k}_\perp^2}{x(1-x)}$.
Similarly in impact space the relevant variable for a two-parton state is  $\zeta^2= x(1-x)\mbf{b}_\perp^2$.
Thus, to first approximation  LF dynamics  depend only on the boost invariant variable
$M_n$ or $\zeta,$
and hadronic properties are encoded in the hadronic mode $\phi(\zeta)$ from the relation (\ref{psiphi})
\begin{equation}  \label{eq:psiphi1}
\psi(x,\zeta, \varphi) = e^{i M \varphi} X(x) \frac{\phi(\zeta)}{\sqrt{2 \pi \zeta}},
\end{equation}
where  the angular dependence $\varphi$, the longitudinal, $X(x)$, and transverse mode $\phi(\zeta)$
have been factored out. The LFWF $\phi(\zeta)$ has normalization
$ \langle\phi\vert\phi\rangle = \int \! d \zeta \,
 \vert \langle \zeta \vert \phi\rangle\vert^2 = 1$. 

We can write the Laplacian operator in (\ref{eq:Mb}) in circular cylindrical coordinates $(\zeta, \varphi)$
and factor out the angular dependence of the
modes in terms of the $SO(2)$ Casimir representation $L^2$ of orbital angular momentum in the
transverse plane. Using  (\ref{eq:psiphi1}) we find~\cite{deTeramond:2008ht}
\begin{equation} \label{eq:KV}
M^2   =  \int \! d\zeta \, \phi^*(\zeta) \sqrt{\zeta}
\left( -\frac{d^2}{d\zeta^2} -\frac{1}{\zeta} \frac{d}{d\zeta}
+ \frac{L^2}{\zeta^2}\right)
\frac{\phi(\zeta)}{\sqrt{\zeta}}   \\
+ \int \! d\zeta \, \phi^*(\zeta) \, U(\zeta)  \, \phi(\zeta) ,
\end{equation}
where all the complexity of the interaction terms in the QCD Lagrangian is summed up in the effective potential $U(\zeta)$.
The light-front eigenvalue equation $H_{LF} \vert \phi \rangle = M^2 \vert \phi \rangle$
is thus a LF wave equation for $\phi$
\begin{equation} \label{LFWE}
\left(-\frac{d^2}{d\zeta^2}
- \frac{1 - 4L^2}{4\zeta^2} + U(\zeta) \right)
\phi(\zeta) = M^2 \phi(\zeta),
\end{equation}
an effective single-variable light-front Schr\"odinger equation which is
relativistic, covariant and analytically tractable. Using (\ref{eq:Mb}) one can readily
generalize the equations to allow for the kinetic energy of massive
quarks.~\cite{Brodsky:2008pg}  In this case, however,
the longitudinal mode $X(x)$ does not decouple from the effective LF bound-state equations.
The mapping  of transition matrix elements
 for arbitrary values of the momentum transfer described in Sec. \ref{EMFF} gives $X(x) = \sqrt{x(1-x)}$~\cite{Brodsky:2006uqa,Brodsky:2007hb,Brodsky:2008pf} in the limit of zero quark masses.

We now compare (\ref{LFWE}) with the wave equation in AdS$_{d+1}$ space for a spin-$J$ mode $\Phi_J$, $\Phi_J= \Phi_{\mu_1 \mu_2 \cdots \mu_J}$, with all the polarization indices
along the physical 3 + 1 coordinates~\cite{deTeramond:2008ht, deTeramond:2010ge}
\footnote{A detailed discussion of higher integer and half-integer spin wave equations  in modified AdS spaces
will be given in~\cite{BDdT:2011}. See also the discussion in Ref. \cite{Gutsche:2011vb}.}
\begin{equation} \label{WeJ}
\left[-\frac{ z^{d-1 -2 J}}{e^{\varphi(z)}}   \partial_z \left(\frac{e^{\varphi(z)}}{z^{d-1 - 2 J}} \partial_z\right)
+ \left(\frac{\mu R}{z}\right)^2\right] \Phi_{\mu_1 \mu_2 \cdots \mu_J} = M^2 \Phi_{\mu_1 \mu_2 \cdots \mu_J}.
\end{equation}

Upon the substitution $z \! \to\! \zeta$  and
$\phi_J(\zeta)   = \left(\zeta/R\right)^{-3/2 + J} e^{\varphi(z)/2} \, \Phi_J(\zeta)$
in (\ref{WeJ}), we find for $d=4$ the QCD light-front wave equation (\ref{LFWE}) with
the effective potential~\cite{deTeramond:2010ge}
\begin{equation} \label{U}
U(\zeta) = \half \varphi''(z) +\frac{1}{4} \varphi'(z)^2  + \frac{2J - 3}{2 z} \varphi'(z) ,
\end{equation}
where the fifth dimensional mass $\mu$ is not a free parameter but scales as $(\mu R)^2 = - (2-J)^2 + L^2$. If $L^2 \ge 0$ the LF Hamiltonian is positive definite
 $\langle \phi \vert H_{LF} \vert \phi \rangle \ge 0$ and thus $ M^2 \ge 0$.
 If $L^2 < 0$ the bound state equation is unbounded from below. The critical value corresponds to $L=0$.
 The quantum mechanical stability $L^2 >0$ for $J=0$ is thus equivalent to the
 Breitenlohner-Freedman stability bound in AdS.~\cite{Breitenlohner:1982jf}
 The AdS equations
correspond to the kinetic energy terms of  the partons inside a
hadron, whereas the interaction terms build confinement.

In the hard-wall model one has $U(z)=0$; confinement is introduced by requiring the wavefunction to vanish at $z=z_0 \equiv 1/\Lambda_{\rm QCD}.$~\cite{Polchinski:2001tt}
In the case of the soft-wall model,~\cite{Karch:2006pv}  the potential arises from a  ``dilaton'' modification of the AdS metric; it  has the form of a harmonic oscillator. For the confining  positive-sign dilaton background $\exp(+ \kappa^2 z^2)$~\cite {deTeramond:2009xk, Andreev:2006ct}  we find the effective potential
$U(z) = \kappa^4 z^2 + 2 \kappa^2(L+S-1)$. The resulting mass spectra  for mesons  at zero quark mass is
${\cal M}^2 = 4 \kappa^2 (n + L +S/2)$.

\begin{figure}[h]
\begin{center}
\includegraphics[width=7.2cm]{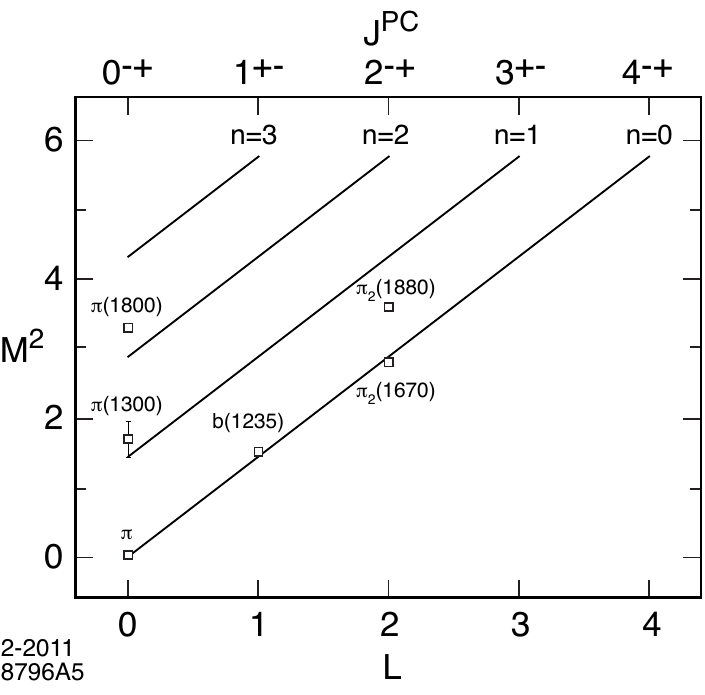}  \hspace{10pt}
\includegraphics[width=7.2cm]{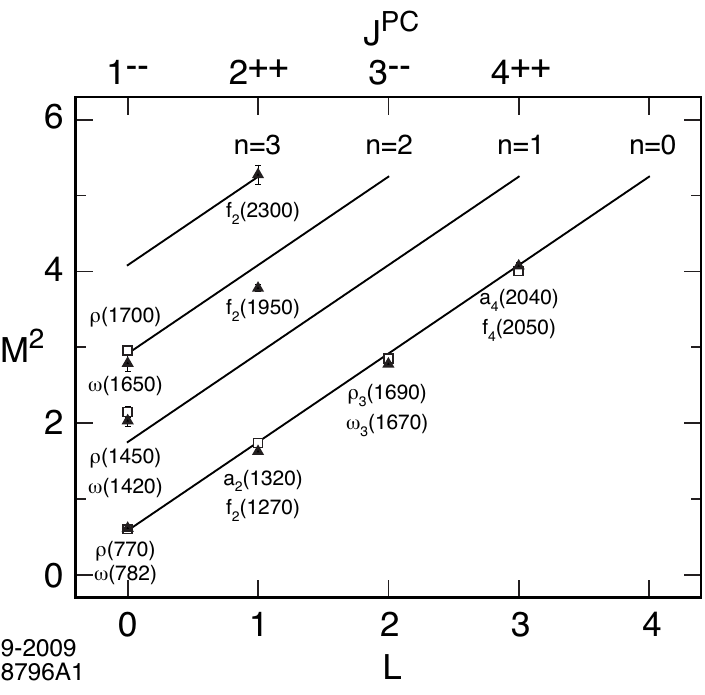}
 \caption{Parent and daughter Regge trajectories for (a) the $\pi$-meson family with
$\kappa= 0.6$ GeV; and (b) the  $I\!=\!1$ $\rho$-meson
 and $I\!=\!0$  $\omega$-meson families with $\kappa= 0.54$ GeV. Only confirmed PDG states~\cite{Amsler:2008xx} are shown.}
\label{pionspec}
\end{center}
\end{figure}

\begin{figure}[h]
\begin{center}
\includegraphics[width=10cm]{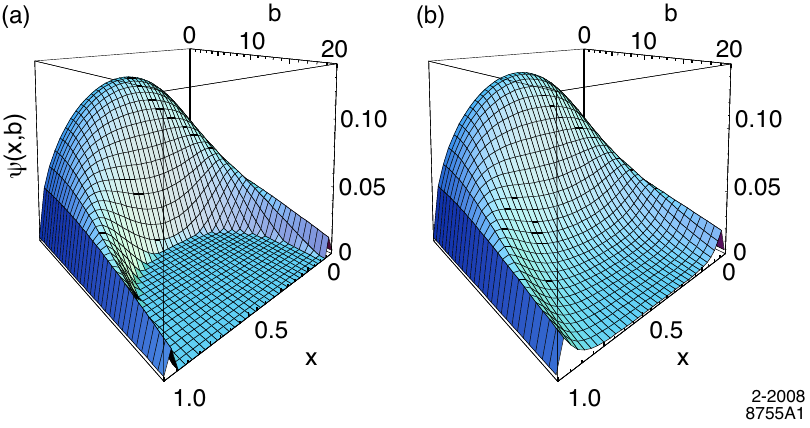}
 \caption{Pion light-front wavefunction $\psi_\pi(x, \mbf{b}_\perp$) from light-front holographic mapping: (a) hard-wall and (b) soft-wall  models.}
\label{LFWF}
\end{center}
\end{figure}

The spectral predictions for  light meson and vector meson states are compared with experimental data
in Fig. \ref{pionspec} for the positive-sign dilaton model discussed here.
The corresponding wavefunctions for the hard and soft-wall models (see Fig.  \ref{LFWF})
display confinement at large interquark
separation and conformal symmetry at short distances, reproducing the dimensional counting rules~\cite{Brodsky:1973kr,Matveev:1973ra} for hard exclusive amplitudes.

\section{Baryons in Light-Front Holography}

For baryons, the light-front wave equation is a linear equation
determined by the LF transformation properties of spin 1/2 states. A linear confining potential
$U(\zeta) \sim \kappa^2 \zeta$ in the LF Dirac
equation leads to linear Regge trajectories.~\cite{Brodsky:2008pg}   For fermionic modes the light-front matrix
Hamiltonian eigenvalue equation $D_{LF} \vert \psi \rangle = M \vert \psi \rangle$, $H_{LF} = D_{LF}^2$,
in a $2 \times 2$ spinor  component
representation is equivalent to the system of coupled linear equations
\begin{eqnarray} \label{eq:LFDirac} \nonumber
- \frac{d}{d\zeta} \psi_- -\frac{\nu+\half}{\zeta}\psi_-
- \kappa^2 \zeta \psi_-&=&
M \psi_+, \\ \label{eq:cD2k}
  \frac{d}{d\zeta} \psi_+ -\frac{\nu+\half}{\zeta}\psi_+
- \kappa^2 \zeta \psi_+ &=&
M \psi_-.
\end{eqnarray}
with eigenfunctions
\begin{eqnarray} \nonumber
\psi_+(\zeta) &\sim& z^{\frac{1}{2} + \nu} e^{-\kappa^2 \zeta^2/2}
  L_n^\nu(\kappa^2 \zeta^2) ,\\
\psi_-(\zeta) &\sim&  z^{\frac{3}{2} + \nu} e^{-\kappa^2 \zeta^2/2}
 L_n^{\nu+1}(\kappa^2 \zeta^2),
\end{eqnarray}
and  eigenvalues
\begin{equation}
M^2 = 4 \kappa^2 (n + \nu + 1) .
\end{equation}

The baryon interpolating operator
$ \mathcal{O}_{3 + L} =  \psi D_{\{\ell_1} \dots
 D_{\ell_q } \psi D_{\ell_{q+1}} \dots
 D_{\ell_m\}} \psi$,  $L = \sum_{i=1}^m \ell_i$, is a twist 3,  dimension $9/2 + L$
 operator
 with scaling behavior given by its
 twist-dimension $3 + L$. We thus require $\nu = L+1$ to match the short distance scaling behavior. Higher spin fermionic modes are obtained by shifting dimensions for the fields as in the bosonic case.
Thus, as in the meson sector,  the increase  in the
mass squared for  higher baryonic states is
$\Delta n = 4 \kappa^2$, $\Delta L = 4 \kappa^2$ and $\Delta S = 2 \kappa^2,$
relative to the lowest ground state,  the proton.
Since our starting point to find the bound state equation of motion for baryons is the light-front, we fix the overall energy scale identical for mesons and baryons by imposing chiral symmetry to the pion~\cite{deTeramond:2010we} in the LF Hamiltonian equations. By contrast, if we start with a five-dimensional action for a scalar field in presence of a positive sign dilaton, the pion is automatically massless.

\begin{figure}[!]
\begin{center}
\includegraphics[angle=0,width=12.5cm]{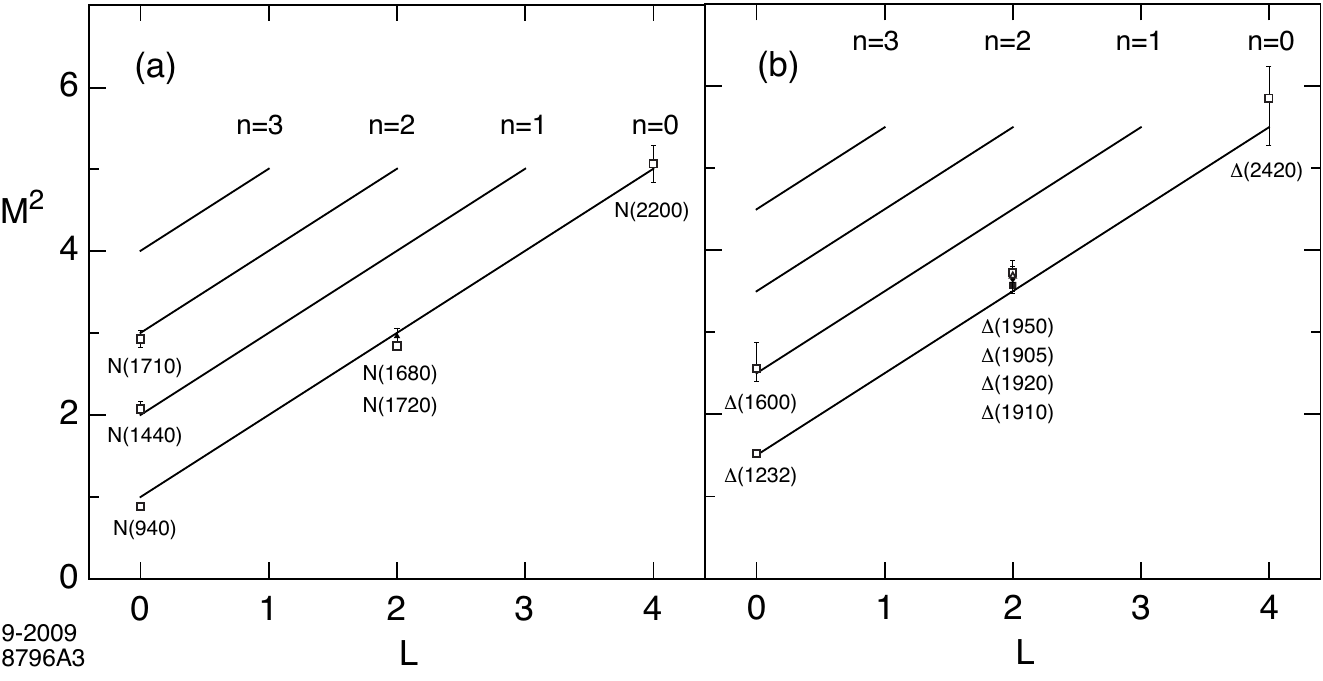}
\caption{{Positive parity Regge trajectories for  the  $N$ and $\Delta$ baryon families for $\kappa= 0.5$ GeV.}}
\label{Baryons}
\end{center}
\end{figure}

The predictions for the positive parity light baryons are shown in Fig. \ref{Baryons}.
As for the predictions for mesons in Fig. \ref{pionspec}, only confirmed PDG~\cite{Amsler:2008xx} states are shown.
The Roper state $N(1440)$ and the $N(1710)$ are well accounted for in this model as the first  and second radial
states. Likewise the $\Delta(1660)$ corresponds to the first radial state of the $\Delta$ family. The model is  successful in explaining the important parity degeneracy observed in the light baryon spectrum, such as the $L\! =\!2$, $N(1680)\!-\!N(1720)$ degenerate pair and the $L=2$, $\Delta(1905), \Delta(1910), \Delta(1920), \Delta(1950)$ states which are degenerate
within error bars. Parity degeneracy of baryons is also a property of the hard wall model, but radial states are not well described in this model.~\cite{deTeramond:2005su}
For other calculations of the baryonic spectrum in the framework of AdS/QCD, see Refs.~\cite{Hong:2006ta,  Forkel:2007cm, Nawa:2008xr,  Forkel:2008un, Ahn:2009px,  Zhang:2010bn, Kirchbach:2010dm}.

An important feature of light-front holography is that it predicts the same multiplicity of states for mesons
and baryons as it is observed experimentally.~\cite{Klempt:2007cp} This remarkable property could have a simple explanation in the cluster decomposition of the
holographic variable $\zeta$ (\ref{zeta}), which labels a system of partons as an active quark plus a system on $n-1$ spectators. From this perspective a baryon with $n=3$ looks in light-front holography as a quark-diquark system.

Nonzero quark masses are naturally incorporated into the AdS/LF predictions~\cite{Brodsky:2008pg, Branz:2010ub} by including them explicitly in the LF kinetic energy  $\sum_i ( {\mbf{k}^2_{\perp i} + m_i^2})/{x_i}$. Given the nonperturbative LFWFs one can predict many interesting phenomenological quantities such as heavy quark decays, generalized parton distributions and parton structure functions.  The AdS/QCD model is semiclassical, and thus it only predicts the lowest valence Fock state structure of the hadron LFWF. One can systematically improve the holographic approximation by
diagonalizing the QCD LF Hamiltonian on the AdS/QCD basis~\cite{Vary:2009gt}, or by using the Lippmann-Schwinger equations.
The action of the non-diagonal terms
in the QCD interaction Hamiltonian also generates the form of the higher
Fock state structure of hadronic LFWFs.  In
contrast with the original AdS/CFT correspondence, the large $N_C$
limit is not required to connect light-front QCD to
an effective dual gravity approximation.

\section {Vacuum Effects and Light-Front Quantization\label{VacuumEffects}}

The LF vacuum is remarkably simple in light-front quantization because of the restriction $k^+ \ge 0.$   For example in QED,  vacuum graphs such as $e^+ e^- \gamma $  associated with the zero-point energy do not arise. In the Higgs theory, the usual Higgs vacuum expectation value is replaced with a $k^+=0$ zero mode;~\cite{Srivastava:2002mw} however, the resulting phenomenology is identical to the standard analysis.

Hadronic condensates play an important role in quantum chromodynamics (QCD).
Conventionally, these condensates are considered to be properties
of the QCD vacuum and hence to be constant throughout space-time.
Recently a new perspective on the nature of QCD
condensates $\langle \bar q q \rangle$ and $\langle
G_{\mu\nu}G^{\mu\nu}\rangle$, particularly where they have spatial and temporal
support,
has been presented.~\cite{Brodsky:2008be,Brodsky:2008xu,Brodsky:2009zd}
Their spatial support is restricted to the interior
of hadrons, since these condensates arise due to the interactions of quarks and
gluons which are confined within hadrons. For example, consider a meson consisting of a light quark $q$ bound to a heavy
antiquark, such as a $B$ meson.  One can analyze the propagation of the light
$q$ in the background field of the heavy $\bar b$ quark.  Solving the
Dyson-Schwinger equation for the light quark one obtains a nonzero dynamical
mass and, via the connection mentioned above, hence a nonzero value of the
condensate $\langle \bar q q \rangle$.  But this is not a true vacuum
expectation value; instead, it is the matrix element of the operator $\bar q q$
in the background field of the $\bar b$ quark.  The change in the (dynamical)
mass of the light quark in this bound state is somewhat reminiscent of the
energy shift of an electron in the Lamb shift, in that both are consequences of
the fermion being in a bound state rather than propagating freely.
Similarly, it is important to use the equations of motion for confined quarks
and gluon fields when analyzing current correlators in QCD, not free
propagators, as has often been done in traditional analyses of operator
products.  Since after a $q \bar q$ pair is created, the distance between the
quark and antiquark cannot get arbitrarily great, one cannot create a quark
condensate which has uniform extent throughout the universe.
As a result, it is argued in Refs. ~\cite{Brodsky:2008be,Brodsky:2008xu,Brodsky:2009zd}    that the 45 orders of magnitude conflict of QCD with the observed value of the cosmological condensate is removed.
A new perspective on the nature of quark and gluon condensates in
quantum chromodynamics is thus obtained:~\cite{Brodsky:2008be,Brodsky:2008xu,Brodsky:2009zd}  the spatial support of QCD condensates
is restricted to the interior of hadrons, since they arise due to the
interactions of confined quarks and gluons.  In the LF theory, the condensate physics is replaced by the dynamics of higher non-valence Fock states as shown by Casher and Susskind.~\cite{Casher:1974xd}  In particular, chiral symmetry is broken in a limited domain of size $1/ m_\pi$,  in analogy to the limited physical extent of superconductor phases.  

This novel description  of chiral symmetry breaking  in terms of ``in-hadron condensates"  has also been observed in Bethe-Salpeter studies~\cite{Maris:1997hd,Maris:1997tm}.
The usual argument for a quark vacuum condensate is the Gell-Mann--Oakes--Renner formula:
\begin{equation}
m^2_\pi = -2 m_q {\langle0| \bar q q |0\rangle\over f^2_\pi}.
\end{equation}
However, in the Bethe-Salpeter formalism, where the pion is a $q \bar q$ bound-state, the GMOR relation is replaced by
\begin{equation}
m^2_\pi = - 2 m_q {\langle 0| \bar q \gamma_5  q |\pi \rangle\over f_\pi},
\end{equation}
where $\rho_\pi \equiv - \langle0| \bar q \gamma_5  q |\pi\rangle$  represents a pion decay constant via an an elementary pseudoscalar current. The result is independent of the renormalization scale. In the light-front formalism, this matrix element derives from the $|q \bar q \rangle$ Fock state of the pion with parallel spin-projections $S^z = \pm 1$ and $L^z= \mp 1$, which couples by quark spin-flip to the usual $|q \bar q\rangle$ $S^z=0, L^z=0$ Fock state via the running quark mass. 
This new perspective explains the
results of studies~\cite{Ioffe:2002be,Davier:2007ym,Davier:2008sk} which find no significant signal for the vacuum gluon
condensate.

AdS/QCD also provides  a description of chiral symmetry breaking by
using the propagation of a scalar field $X(z)$
to represent the dynamical running quark mass.
In the hard wall model the solution has the form~\cite{Erlich:2005qh,DaRold:2005zs} $X(z) = a_1 z+ a_2 z^3$, where $a_1$ is
proportional to the current-quark mass. The coefficient $a_2$ scales as
$\Lambda^3_{QCD}$ and is the analog of $\langle \bar q q \rangle$; however,
since the quark is a color nonsinglet, the propagation of $X(z),$ and thus the
domain of the quark condensate, is limited to the region of color confinement.
Furthermore the effect of the $a_2$ term
varies within the hadron, as characteristic of an in-hadron condensate.
The AdS/QCD picture of condensates with spatial support restricted to hadrons
is also in general agreement with results from chiral bag
models,~\cite{Chodos:1975ix,Brown:1979ui,Hosaka:1996ee}
which modify the original MIT bag by coupling a pion field to the surface of
the bag in a chirally invariant manner.

\section{Hadronization at the Amplitude Level \label{hadronization}}

\begin{figure}[!]
\centering
\includegraphics[width=10cm]{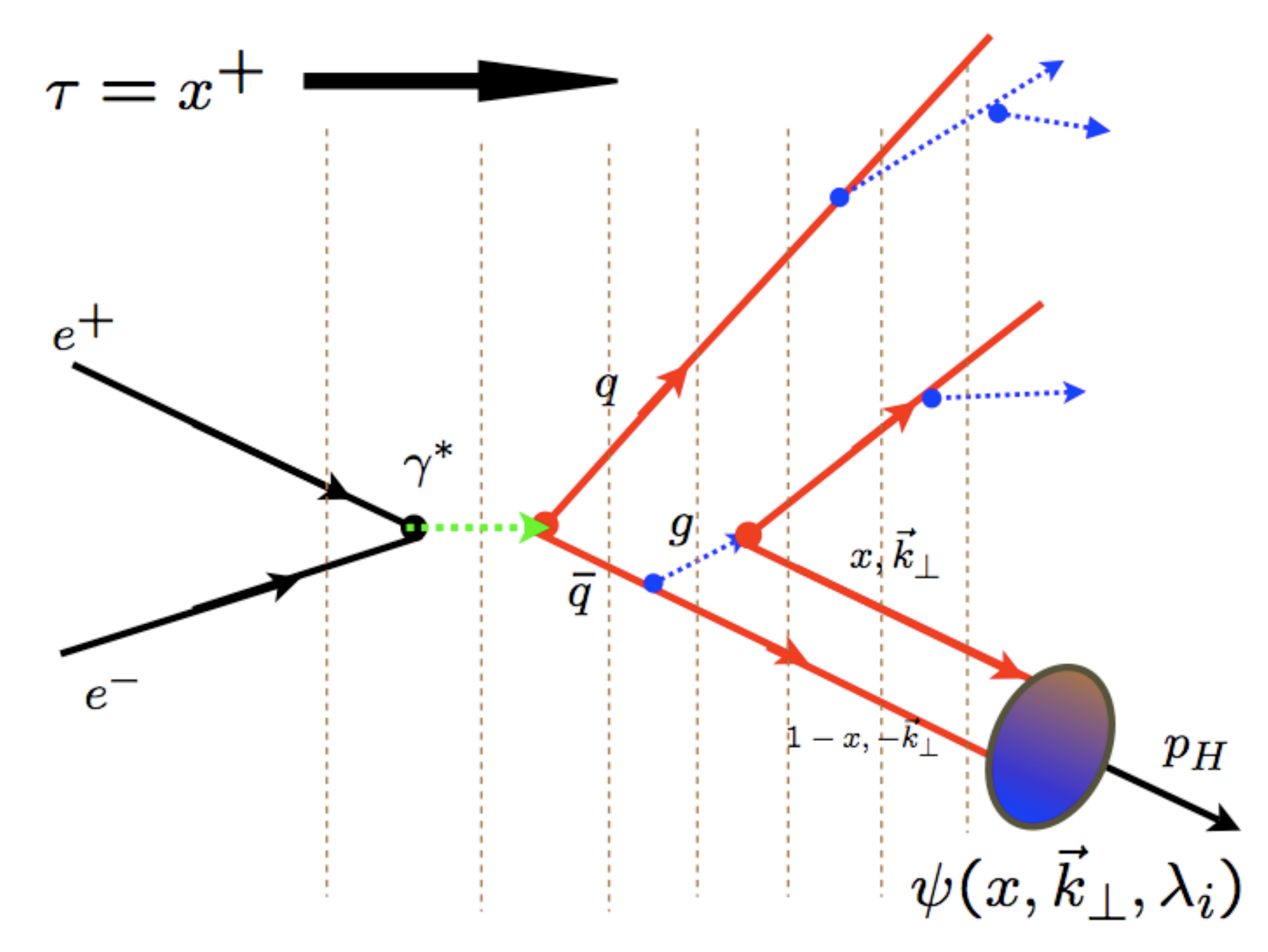}
  \caption{Illustration of an event amplitude generator for $e^+ e^- \to \gamma^* \to X$ for
  hadronization processes at the amplitude level. Capture occurs if the quarks try to go beyond the confinement distance;  i.e, if
  $\zeta^2 = x(1-x) \mbf{b}_\perp^2 > 1/ \Lambda_{\rm QCD}^2$
   in the AdS/QCD hard-wall model of confinement. The corresponding condition in momentum space is
  $M^2 = \frac{\mbf{k}_\perp^2}{x(1-x)} \lesssim \Lambda_{\rm QCD}^2$.}
\label{hadroniz}
\end{figure}

The conversion of quark and gluon partons  to hadrons is usually discussed in terms  of on-shell hard-scattering cross sections convoluted with {\it ad hoc} probability distributions.
The LF Hamiltonian formulation of quantum field theory provides a natural formalism to compute
hadronization at the amplitude level.~\cite{Brodsky:2008tk}  In this case, one uses light-front time-ordered perturbation theory for the QCD light-front Hamiltonian to generate the off-shell  quark and gluon T-matrix helicity amplitude  using the LF generalization of the Lippmann-Schwinger formalism:
\begin{equation}
T ^{LF}=
{H^{LF}_I }  \\ +
{H^{LF}_I }{1 \over M^2_{\rm Initial} -  M^2_{\rm intermediate} + i \epsilon} {H^{LF}_I }
+ \cdots
\end{equation}
Here   $M^2_{\rm intermediate}  = \sum^N_{i=1} {(\mbf{k}^2_{\perp i} + m^2_i )/x_i}$ is the invariant mass squared of the intermediate state and ${H^{LF}_I }$ is the set of interactions of the QCD LF Hamiltonian in the ghost-free light-cone gauge.~\cite{Brodsky:1997de}
The $T^{LF}$ matrix element is
evaluated between the out and in eigenstates of $H^{QCD}_{LF}$.   The event amplitude generator is illustrated for $e^+ e^- \to \gamma^* \to X$ in Fig. \ref{hadroniz}.

The LFWFs of AdS/QCD can be used as the interpolating amplitudes between the off-shell quark and gluons and the bound-state hadrons.
Specifically,
if at any stage a set of  color-singlet partons has  light-front kinetic energy
$\sum_i {\mbf{k}^2_{\perp i}/ x_i} \!  < \! \Lambda^2_{\rm QCD}$, then one coalesces the virtual partons into a hadron state using the AdS/QCD LFWFs.   This provides a specific scheme for determining the factorization scale which  matches perturbative and nonperturbative physics.

This scheme has a number of  important computational advantages:

(a) Since propagation in LF Hamiltonian theory only proceeds as $\tau$ increases, all particles  propagate as forward-moving partons with $k^+_i \ge 0$.  There are thus relatively few contributing
 $\tau$-ordered diagrams.

(b) The computation
implementation can be highly efficient: an amplitude of order $g^n$ for a given process only needs to be computed once.  In fact, each non-interacting cluster within $T^{LF}$ has a numerator which is process independent; only the LF denominators depend on the context of the process.  This method has recently been used by   L.~Motyka and A.~M.~Stasto~\cite{Motyka:2009gi}
to compute gluonic scattering amplitudes in QCD.

(c) Each amplitude can be renormalized using the ``alternate denominator'' counterterm method, rendering all amplitudes UV finite.~\cite{Brodsky:1973kb}

(d) The renormalization scale in a given renormalization scheme  can be determined for each skeleton graph even if there are multiple physical scales.

(e) The $T^{LF}$ matrix computation allows for the effects of initial and final state interactions of the active and spectator partons. This allows for leading-twist phenomena such as diffractive DIS, the Sivers spin asymmetry and the breakdown of the PQCD Lam-Tung relation in Drell-Yan processes.

(f)  ERBL and DGLAP evolution are naturally incorporated, including the quenching of  DGLAP evolution  at large $x_i$ where the partons are far off-shell.

(g) Color confinement can be incorporated at every stage by limiting the maximum wavelength of the propagating quark and gluons.

(h) This method retains the quantum mechanical information in hadronic production amplitudes which underlie Bose-Einstein correlations and other aspects of the spin-statistics theorem.
Thus Einstein-Podolsky-Rosen QM correlations are maintained even between far-separated hadrons and  clusters.

A similar off-shell T-matrix approach was used to predict antihydrogen formation from virtual positron--antiproton states produced in $\bar p A$
collisions.~\cite{Munger:1993kq}

\section{Dynamical Effects of Rescattering \label{rescat}}

Initial-state and final-state rescatterings,
neglected in the parton model, have a profound effect in QCD hard-scattering reactions,
predicting single-spin asymmetries,~\cite{Brodsky:2002cx,Collins:2002kn} diffractive deep lepton-hadron inelastic scattering,~\cite{Brodsky:2002ue} the breakdown of
the Lam Tung relation in Drell-Yan reactions,~\cite{Boer:2002ju} nor nuclear shadowing and non-universal
antishadowing~\cite{Brodsky:2004qa}---leading-twist physics which is not incorporated in
the light-front wavefunctions of the target computed in isolation.
It is thus important to distinguish~\cite{Brodsky:2008xe} ``static'' or ``stationary'' structure functions which are computed directly from the LFWFs of the target  from the ``dynamic'' empirical structure functions which take into account rescattering of the struck quark.   Since they derive from the LF eigenfunctions of the target hadron, the static structure functions have a probabilistic interpretation.  The wavefunction of a stable eigenstate is real; thus the static structure functions cannot describe diffractive deep inelastic scattering nor the single-spin asymmetries since such phenomena involves the complex phase structure of the $\gamma^* p $ amplitude.
One can augment the light-front wavefunctions with a gauge link corresponding to an external field
created by the virtual photon $q \bar q$ pair
current,~\cite{Belitsky:2002sm,Collins:2004nx} but such a gauge link is
process dependent,~\cite{Collins:2002kn} so the resulting augmented
wavefunctions are not universal.~\cite{Brodsky:2002ue,Belitsky:2002sm,Collins:2003fm}

It should be emphasized
that the shadowing of nuclear structure functions is due to the
destructive interference between multi-nucleon amplitudes involving
diffractive DIS and on-shell intermediate states with a complex
phase.  The physics of rescattering and shadowing is thus not
included in the nuclear light-front wavefunctions, and a
probabilistic interpretation of the nuclear DIS cross section is
precluded.
In addition, one finds that antishadowing in deep inelastic lepton-nucleus scattering is
not universal,~\cite{Brodsky:2004qa}
but depends on the flavor of each quark and antiquark struck by the lepton.  Evidence of this phenomena has been reported by
Schienbein {\it et al}.~\cite{Schienbein:2009kk}

\begin{figure}[!]
\centering
\includegraphics[width=13cm]{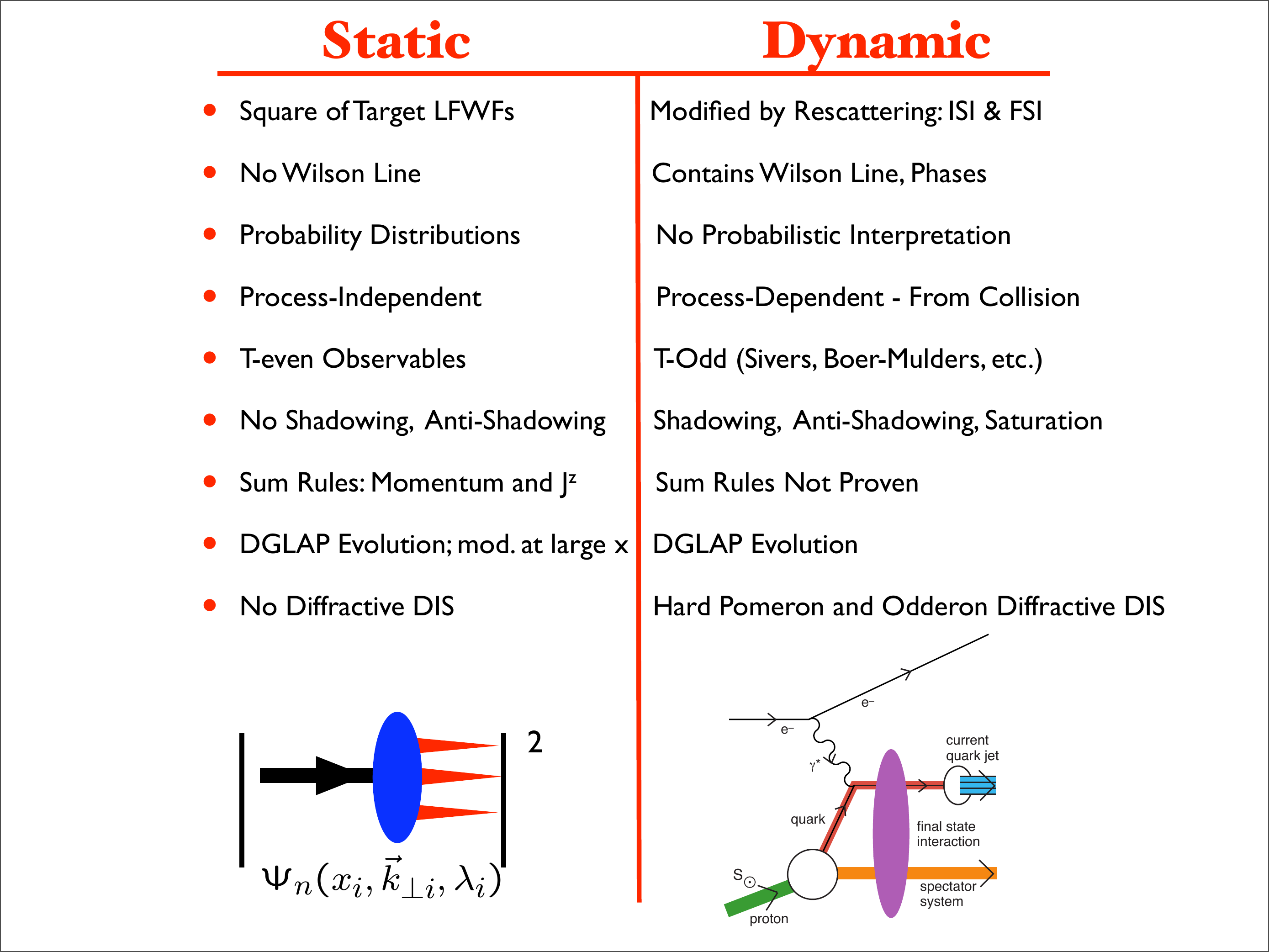}
\caption{Static vs dynamic structure functions}
\label{figstatdyn}
\end{figure}

The distinction
between static structure functions; i.e., the probability distributions  computed from the square of the light-front wavefunctions, versus the nonuniversal dynamic structure functions measured in deep inelastic scattering is summarized in Fig. \ref{figstatdyn}.

\section{Novel Perspectives on QCD from Light-Front Dynamics}

In this section we summarize a number of  topics  where new, and in some cases surprising, perspectives for QCD physics have emerged from the light-front formalism and light-front holography.

\begin{enumerate}

\item  It is natural to assume that the nuclear modifications to the structure functions measured in deep inelastic lepton-nucleus and neutrino-nucleus interactions are identical;  however, the Gribov-Glauber theory predicts that the antishadowing of nuclear structure functions is not  universal, but depends on the quantum numbers of each struck quark and antiquark.~\cite{Brodsky:2004qa}  This observation can explain the recent analysis of Schienbein {\it et al.},~\cite{Schienbein:2008ay} which shows that the NuTeV measurements of nuclear structure functions obtained from neutrino  charged current reactions differ significantly from the distributions measured in deep inelastic electron and muon scattering.

\item The effects of final-state interactions of the scattered quark  in deep inelastic scattering  have been traditionally assumed to be power-law suppressed.  In fact,  the final-state gluonic interactions of the scattered quark lead to a  $T$-odd non-zero spin correlation of
the lepton-quark scattering plane with the polarization of the target proton.~\cite{Brodsky:2002cx}  This  leading-twist ``Sivers effect''  is non-universal since QCD predicts an opposite-sign
correlation~\cite{Collins:2002kn,Brodsky:2002rv} in Drell-Yan reactions, due to the initial-state interactions of the annihilating antiquark.
The final-state interactions of the struck quark with the spectators~\cite{Brodsky:2002ue}  also lead to diffractive events in deep inelastic scattering (DIS) at leading twist,  such as $\ell p \to \ell^\prime p^\prime X ,$
where the proton remains intact and isolated in rapidity;    in fact, approximately 10 \% of the deep inelastic lepton-proton scattering events observed at HERA are
diffractive.~\cite{Adloff:1997sc, Breitweg:1998gc} The presence of a rapidity gap
between the target and diffractive system requires that the target
remnant emerges in a color-singlet state; this is made possible in
any gauge by the soft rescattering incorporated in the Wilson line or by augmented light-front wavefunctions.

\item It is usually assumed --  following the intuition of the parton model -- that the structure functions  measured in deep inelastic scattering can be computed in the Bjorken-scaling leading-twist limit from the absolute square of the light-front wavefunctions, summed over all Fock states.  In fact,  dynamical effects, such as the Sivers spin correlation and diffractive deep inelastic lepton scattering due to final-state gluon interactions,  contribute to the experimentally observed DIS cross sections.
Diffractive events also lead to the interference of two-step and one-step processes in nuclei which in turn, via the Gribov-Glauber theory, lead to the shadowing and the antishadowing of the deep inelastic nuclear structure functions;~\cite{Brodsky:2004qa}  such phenomena are not included in the light-front wavefunctions of the nuclear eigenstate.
This leads to an important  distinction between ``dynamical''  vs. ``static''  (wavefunction-specific) structure functions.~\cite{Brodsky:2009dv}

\item
As  noted by Collins and Qiu,~\cite{Collins:2007nk} the traditional factorization formalism of perturbative QCD  fails in detail for many hard inclusive reactions because of initial- and final-state interactions.  For example, if both the
quark and antiquark in the Drell-Yan subprocess
$q \bar q \to  \mu^+ \mu^-$ interact with the spectators of the
other  hadron, then one predicts a $\cos 2\phi \sin^2 \theta$ planar correlation in unpolarized Drell-Yan
reactions.~\cite{Boer:2002ju}  This ``double Boer-Mulders effect" can account for the large $\cos 2 \phi$ correlation and the corresponding violation~\cite{Boer:2002ju, Boer:1999mm} of the Lam Tung relation for Drell-Yan processes observed by the NA10 collaboration.
An important signal for factorization breakdown at the LHC  will be the observation of a $\cos 2 \phi$ planar correlation in dijet production.

\item	  It is conventional to assume that the charm and bottom quarks in the proton structure functions  only arise from gluon splitting $g \to Q \bar Q.$  In fact, the proton light-front wavefunction contains {\it ab initio } intrinsic heavy quark Fock state components such as $\vert uud c \bar c\rangle$.~\cite{Brodsky:1980pb,Brodsky:1984nx,Harris:1995jx,Franz:2000ee}  The intrinsic heavy quarks carry most of the proton's momentum since this minimizes the off-shellness of the state. The heavy quark pair $Q \bar Q$ in the intrinsic Fock state  is primarily a color-octet,  and the ratio of intrinsic charm to intrinsic bottom scales as $m_c^2/m_b^2 \simeq 1/10,$ as can easily be seen from the operator product expansion in non-Abelian QCD.   Intrinsic charm and bottom explain the origin of high $x_F$ open-charm and open-bottom hadron production, as well as the single and double $J/\psi$ hadroproduction cross sections observed at high $x_F$.   The factorization-breaking nuclear $A^\alpha(x_F)$ dependence  of hadronic $J/\psi$ production cross sections is also explained.  A novel mechanism for inclusive and diffractive
Higgs production $pp \to p p H  $, in which the Higgs boson carries a significant fraction of the projectile proton momentum, is discussed in Ref.~\cite{Brodsky:2006wb}.  The production
mechanism is based on the subprocess $(Q \bar Q) g \to H $ where the $Q \bar Q$ in the $\vert uud Q \bar Q \rangle$ intrinsic heavy quark Fock state of the colliding proton has approximately
$80\%$ of the projectile protons momentum.

\item It is normally assumed that high transverse momentum hadrons in inclusive high energy hadronic collisions,  such as $ p p \to H X$,  can only arise  from jet fragmentation.
A  fundamental test of leading-twist QCD predictions in high transverse momentum hadronic reactions is the measurement of the power-law
fall-off of the inclusive cross section~\cite{Sivers:1975dg}
${E d \sigma/d^3p}(A B \to C X) ={ F(\theta_{cm}, x_T)/ p_T^{n_{eff}} } $ at fixed $x_T = 2 p_T/\sqrt s$
and fixed $\theta_{CM},$ where $n_{eff} \sim 4 + \delta$. Here $\delta  =  {\cal O}(1)$ is the correction to the conformal prediction arising
from the QCD running coupling and the DGLAP evolution of the input parton distribution and fragmentation functions.~\cite{Brodsky:2005fz,Arleo:2009ch,Arleo:2010yg}
The usual expectation is that leading-twist subprocesses will dominate measurements of high $p_T$ hadron production at RHIC and Tevatron energies. In fact, the  data for isolated photon production $ p p \to \gamma_{\rm direct} X,$ as well as jet production, agree
well with the  leading-twist scaling prediction $n_{eff}  \simeq 4.5$.~\cite{Arleo:2009ch}
However,   measurements  of  $n_{eff} $ for hadron production  are not consistent with the leading twist predictions.
Striking
deviations from the leading-twist predictions were also observed at lower energy at the ISR and  Fermilab fixed-target experiments.~\cite{Sivers:1975dg,Cronin:1973fd,Antreasyan:1978cw}
In fact, a significant fraction of high $p^H_\perp$ isolated hadrons can emerge
directly from hard higher-twist subprocess~\cite{Arleo:2009ch,Arleo:2010yg} even at the LHC.  The direct production of hadrons can explain~\cite{Brodsky:2008qp} the remarkable ``baryon anomaly" observed at RHIC:  the ratio of baryons to mesons at high $p^H_\perp$,  as well as the power-law fall-off $1/ p_\perp^n$ at fixed $x_\perp = 2 p_\perp/\sqrt s, $ both  increase with centrality,~\cite{Adler:2003kg} opposite to the usual expectation that protons should suffer more energy loss in the nuclear medium than mesons.
The high values $n_{eff}$ with $x_T$ seen in the data  indicate the presence of an array of higher-twist processes, including subprocesses where the hadron enters directly, rather than through jet fragmentation.~\cite{Blankenbecler:1975ct}

\item	It is often stated that the renormalization scale of the QCD running coupling $\alpha_s(\mu^2_R) $  cannot be fixed, and thus it has to be chosen in an {\it ad hoc} fashion.  In fact, as in QED, the scale can be fixed unambiguously by shifting $\mu_R$  so that all terms associated with the QCD $\beta$ function vanish.  In general, each set of skeleton diagrams has its respective scale. The result series is equivalent to the perturbative expansion of an equivalent conformal theory;  it is thus scheme independent and independent of the choice of the initial renormalization scale ${\mu_R}_0$, thus satisfying Callan-Symanzik invariance.    This is the ``principle of maximal conformality"~\cite{Brodsky:2011ig} - the principle which underlies the BLM scale setting method.
Unlike heuristic scale-setting procedures,  the BLM/PMC method~\cite{Brodsky:1982gc} gives results which are independent of the choice of renormalization scheme, as required by the transitivity property of the renormalization group.   The divergent renormalon terms of order $\alpha_s^n \beta^n n!$ are transferred to the physics of the running coupling.  Furthermore, one retains sensitivity to ``conformal''  effects which arise in higher orders; physical effects which are not associated with QCD  renormalization.  The BLM method also provides scale-fixed,
scheme-independent high precision connections between observables, such as the ``Generalized Crewther Relation'',~\cite{Brodsky:1995tb} as well as other ``Commensurate Scale Relations''.~\cite{Brodsky:1994eh,Brodsky:2000cr}  Clearly the elimination of the renormalization scale ambiguity would greatly improve the precision of QCD predictions and increase the sensitivity of searches for  new physics at the LHC.

\item It is usually assumed that the QCD coupling $\alpha_s(Q^2)$ diverges at $Q^2=0$;  i.e., ``infrared slavery''.  In fact, determinations from lattice gauge theory,  Bethe-Salpeter methods, effective charge measurements, gluon mass phenomena, and AdS/QCD all lead (in their respective scheme) to a finite value of the QCD coupling in the infrared.~\cite{Brodsky:2010ur}  Because of color confinement, the quark and gluon propagators vanish at long
wavelength: $k < \Lambda_{QCD}$, and consequently the quantum loop corrections underlying the  QCD $\beta$-function  decouple in the infrared, and  the coupling  freezes to a finite value at
$Q^2 \to 0$.~\cite{Brodsky:2007hb,Brodsky:2008be}   This observation underlies the use of conformal methods in AdS/QCD.

\item It is conventionally assumed that the vacuum of QCD contains quark $\langle 0 \vert q \bar q \vert 0 \rangle$ and gluon  $\langle 0 \vert  G^{\mu \nu} G_{\mu \nu} \vert 0 \rangle$ vacuum condensates, although the resulting vacuum energy density leads to a $10^{45}$  order-of-magnitude discrepancy with the
measured cosmological constant.~\cite{Brodsky:2009zd}  However, a new perspective has emerged from Bethe-Salpeter and light-front analyses where the QCD condensates are identified as ``in-hadron'' condensates, rather than vacuum entities, but consistent with the Gell Mann-Oakes-Renner  relation.~\cite{Brodsky:2010xf} The ``in-hadron''  condensates become realized as higher Fock states of the hadron when the theory is quantized at fixed light-front time.

\item  In nuclear physics nuclei are composites of nucleons. However, QCD provides a new perspective:~\cite{Brodsky:1976rz,Matveev:1977xt}  six quarks in the fundamental
$3_C$ representation of $SU(3)$ color can combine into five different color-singlet combinations, only one of which corresponds to a proton and
neutron.  The deuteron wavefunction is a proton-neutron bound state at large distances, but as the quark separation becomes smaller,
QCD evolution due to gluon exchange introduces four other ``hidden color'' states into the deuteron
wavefunction.~\cite{Brodsky:1983vf} The normalization of the deuteron form factor observed at large $Q^2$,~\cite{Arnold:1975dd} as well as the
presence of two mass scales in the scaling behavior of the reduced deuteron form factor,~\cite{Brodsky:1976rz} suggest sizable hidden-color
Fock state contributions  in the deuteron
wavefunction.~\cite{Farrar:1991qi}
The hidden-color states of the deuteron can be materialized at the hadron level as   $\Delta^{++}(uuu), \, \Delta^{-}(ddd)$ and other novel quantum
fluctuations of the deuteron. These dual hadronic components become important as one probes the deuteron at short distances, such
as in exclusive reactions at large momentum transfer.  For example, the ratio  ${{d \sigma/ dt}(\gamma d \to \Delta^{++}
\Delta^{-})/{d\sigma/dt}(\gamma d\to n p) }$ is predicted to increase to  a fixed ratio $2:5$ with increasing transverse momentum $p_T.$
Similarly, the Coulomb dissociation of the deuteron into various exclusive channels $e d \to e^\prime + p n, p p \pi^-, \Delta \, \Delta, \cdots$
will have a changing composition as the final-state hadrons are probed at high transverse momentum, reflecting the onset of hidden-color
degrees of freedom.

\item It is usually assumed that the imaginary part of the deeply virtual Compton scattering amplitude is determined at leading twist by  generalized parton distributions, but that the real part has an undetermined  ``$D$-term'' subtraction. In fact, the real part is determined by the  local  two-photon interactions of the quark current in the QCD light-front Hamiltonian.~\cite{Brodsky:2008qu,Brodsky:1971zh}  This contact interaction leads to a real energy-independent contribution to the DVCS amplitude  which is independent of the photon virtuality at fixed  $t$.  The interference of the timelike DVCS amplitude with the Bethe-Heitler amplitude leads to a charge asymmetry in $\gamma p \to \ell^+ \ell^- p$.~\cite{Brodsky:1971zh,Brodsky:1973hm,Brodsky:1972vv}   Such measurements can verify that quarks carry the fundamental electromagnetic current within hadrons.

\end{enumerate}

\section{Conclusions}

A long-sought goal in hadron physics is to find a simple analytic first approximation to QCD analogous to the Schr\"odinger-Coulomb equation of atomic physics.	This problem is particularly challenging since the formalism must be relativistic, color-confining, and consistent with chiral symmetry and its breaking.
As we have reviewed here,
the AdS wave equations, modified by a  non-conformal dilaton background field  which incorporates the confinement interaction, leads, via light-front holography, to a simple
Schr\"odinger-like light-front wave equation.~\cite{deTeramond:2005su,deTeramond:2008ht,Brodsky:2010px}
The result is a single-variable
light-front wave equation in $\zeta$,
the transverse invariant separation between the hadronic constituents,
 with an effective confining potential which determines the eigenspectrum and the light-front wavefunctions of hadrons for general spin and orbital angular momentum.
In fact, $\zeta$ plays the same role in relativistic quantum field theory as the radial coordinate $r$ of non-relativistic
 Schr\"odinger quantum mechanics.
For a positive-dilaton profile, a remarkable one-parameter description of nonperturbative hadron dynamics is obtained.~\cite{deTeramond:2005su,deTeramond:2008ht, Brodsky:2010px} This  model predicts a zero-mass pion for zero-mass quarks and a Regge spectrum of linear trajectories with the same slope in the (leading) orbital angular momentum $L$ of the hadrons and their radial  quantum number $n$. The theory implements chiral symmetry  in a novel way:    the effects of chiral symmetry breaking increase as one goes toward large interquark separation.
In spite of its present limitations, the AdS/QCD approach, together with light-front holography, provides important physical insights into the non-perturbative regime of QCD and its transition to the perturbative domain.

``Light-Front Holography"~\cite{deTeramond:2008ht} also allows one to map
the amplitudes $\phi(z)$ in AdS space  directly to the light-front
wavefunctions defined at fixed light-front time in 3+1 space.
The resulting Lorentz-invariant relativistic light-front wave equations are functions of  the invariant impact variable $\zeta$ which measures the separation of the quark and gluonic constituents within the hadron at equal light-front time.  This correspondence was derived by showing that the Polchinski-Strassler formula~\cite{Polchinski:2001tt}
for form factors in AdS space is equivalent to the Drell-Yan-West~\cite{Drell:1969km, West:1970av} light-front matrix element both for external electromagnetic and gravitational currents.
One then finds an exact mapping between $z$ in AdS space and the invariant impact
separation $\zeta$ in 3+1 space-time.  In the case of two-parton wavefunctions, one has  $\zeta =
\sqrt{x(1-x) b^2_\perp}$, where $b_\perp$ is the usual impact separation conjugate to
$k_\perp$ and $x = k^+/P^+$ is the light-front fraction.  This correspondence  agrees
with the intuition that $z$ is related inversely to the internal relative momentum, but the
relation $z \to \zeta$ is precise and exact. This relation also provides a direct
connection between light-front Hamiltonian equations for bound state systems and the
AdS wave equations. The
hadron eigenstates generally have components with different orbital angular momentum; e.g.,  the proton eigenstate in AdS/QCD with massless quarks has $L=0$ and $L=1$ light-front Fock components with equal probability.

One thus obtains
a semi-classical frame-independent first approximation to the spectra and light-front wavefunctions of meson and baryon light-quark  bound states,  which in turn predicts  the
behavior of the pion and nucleon  form factors.  The identification of the coordinate $z$ in AdS space  with $\zeta$ at fixed light-front time also provides a physical understanding of the dynamics described by AdS/QCD. The  $\zeta$ dependence of the relativistic light-front wave equations determines the off-shell dynamics of the bound states as a function of the invariant mass of the constituents.  The variable $L$, which appears as a parameter in the five-dimensional mass parameter  $\mu R$ in AdS space, is identified as the kinematic orbital angular momentum $L^z$ of the constituents in 3+1 space at fixed light-front time.

Given the light-front wavefunctions, one can compute  a wide range of hadron properties, including decay constants, structure functions, distribution amplitudes and hadronic form factors.   The AdS/QCD light-front wavefunctions also lead to a method for computing the hadronization of quark and gluon jets at the amplitude level.~\cite{Brodsky:2008tk}

The AdS/QCD soft-wall model also predicts the form of the non-perturbative effective coupling $\alpha_s^{AdS}(Q)$ and its $\beta$-function.~\cite{Brodsky:2010ur}
The AdS/QCD model can be systematically improved  by using its complete orthonormal solutions to diagonalize the full QCD light-front Hamiltonian~\cite{Vary:2009gt} or by applying the Lippmann-Schwinger method in order to systematically include the QCD interaction terms.

We have also reviewed some novel features  of QCD,  including
the consequences of confinement for quark and gluon condensates.
The distinction
between static structure functions, such as the probability
distributions  computed from the square of the light-front
wavefunctions, versus dynamical structure functions which include the
effects of rescattering,
has also been emphasized.
We have also discussed the relevance of the light-front Hamiltonian formulation of QCD to describe the
coalescence of quark and gluons into hadrons.~\cite{Brodsky:2008tk}

\section*{Acknowledgments}
Lectures presented by SJB and GdT at the Workshop AdS/CFT and Novel Approaches to Hadron and Heavy Ion Physics, Kavli Institute of Theoretical Physics (KITPC), Beijing, China, October 19 and 20, 2010.    We are grateful to the KITPC for  outstanding hospitality, and we thank our collaborators for many helpful conversations.
This research was supported by the Department of Energy  contract DE--AC02--76SF00515.  SLAC-PUB-14525.

\end{document}